\documentclass[conference]{IEEEtran}
\IEEEoverridecommandlockouts
\usepackage{cite}
\usepackage{amsmath,amssymb,amsfonts}
\usepackage{algorithm}
\usepackage{algpseudocode}
\usepackage{graphicx}
\usepackage{textcomp}
\usepackage{xcolor}
\def\BibTeX{{\rm B\kern-.05em{\sc i\kern-.025em b}\kern-.08em
    T\kern-.1667em\lower.7ex\hbox{E}\kern-.125emX}}
\usepackage{amssymb}
\usepackage{amsmath}
\usepackage{pgfplots}
\usepackage{comment}
\usepackage{multirow}
\usepackage{mathbbol}
\usepackage{caption}
\usepackage{booktabs}
\usepackage{caption}
\usepackage{tikz}
\usetikzlibrary{arrows}
\usepackage[capitalize]{cleveref}
\crefname{section}{Sec.}{Secs.}
\usetikzlibrary{shapes.geometric, arrows}

\usepackage{flushend}

\usepackage{siunitx}

\usepackage[font=scriptsize]{subcaption}
\usepackage[font=footnotesize]{caption}

\usepackage{mathtools}

\usepackage{tikz}
\usepackage{pgfplots}
\usepackage{tikzscale}
\usepackage{tikz-qtree}

\pgfplotsset{compat=newest}
\pgfplotsset{plot coordinates/math parser=false}
\pgfplotsset{every axis/.append style={
                    label style={font=\scriptsize},
                    tick label style={font=\scriptsize},
                    legend style={font=\scriptsize}
                    }}
\usetikzlibrary{plotmarks,shapes,patterns,decorations.pathreplacing,backgrounds,calc,arrows,arrows.meta,spy,matrix,shadows,trees,positioning,fit}
\usepgfplotslibrary{patchplots,groupplots}

\tikzstyle{startstop} = [rectangle, rounded corners, minimum width=2cm, minimum height=0.5cm,text centered, draw=black]
\tikzstyle{io} = [trapezium, trapezium left angle=70, trapezium right angle=110, minimum width=3cm, minimum height=1cm, text centered, draw=black]
\tikzstyle{process} = [rectangle, minimum width=2cm, minimum height=0.5cm, text centered, draw=black, alignb=center]
\tikzstyle{decision} = [ellipse, minimum width=2cm, minimum height=1cm, text centered, draw=black]
\tikzstyle{arrow} = [thick,<->,>=stealth]
\tikzstyle{line} = [thick,>=stealth]
\tikzstyle{darrow} = [thick,<->,>=stealth,dashed]
\tikzstyle{sarrow} = [thick,->,>=stealth]
\tikzstyle{larrow} = [line width=0.1mm,dashdotted,->,>=stealth]

\pgfkeys{/pgf/number format/.cd,1000 sep={\,}} 

\makeatletter
\def\grd@save@target#1{%
  \def\grd@target{#1}}
\def\grd@save@start#1{%
  \def\grd@start{#1}}
\tikzset{
  grid with coordinates/.style={
    to path={%
      \pgfextra{%
        \edef\grd@@target{(\tikztotarget)}%
        \tikz@scan@one@point\grd@save@target\grd@@target\relax
        \edef\grd@@start{(\tikztostart)}%
        \tikz@scan@one@point\grd@save@start\grd@@start\relax
        \draw[minor help lines] (\tikztostart) grid (\tikztotarget);
        \draw[major help lines] (\tikztostart) grid (\tikztotarget);
        \grd@start
        \pgfmathsetmacro{\grd@xa}{\the\pgf@x/1cm}
        \pgfmathsetmacro{\grd@ya}{\the\pgf@y/1cm}
        \grd@target
        \pgfmathsetmacro{\grd@xb}{\the\pgf@x/1cm}
        \pgfmathsetmacro{\grd@yb}{\the\pgf@y/1cm}
        \pgfmathsetmacro{\grd@xc}{\grd@xa + \pgfkeysvalueof{/tikz/grid with coordinates/major step x}}
        \pgfmathsetmacro{\grd@yc}{\grd@ya + \pgfkeysvalueof{/tikz/grid with coordinates/major step y}}
        \foreach \x in {\grd@xa,\grd@xc,...,\grd@xb}
        \node[anchor=north] at (\x,\grd@ya) {\pgfmathprintnumber{\x}};
        \foreach \y in {\grd@ya,\grd@yc,...,\grd@yb}
        \node[anchor=east] at (\grd@xa,\y) {\pgfmathprintnumber{\y}};
      }
    }
  },
  minor help lines/.style={
    help lines,
    gray,
    line cap =round,
    xstep=\pgfkeysvalueof{/tikz/grid with coordinates/minor step x},
    ystep=\pgfkeysvalueof{/tikz/grid with coordinates/minor step y}
  },
  major help lines/.style={
    help lines,
    line cap =round,
    line width=\pgfkeysvalueof{/tikz/grid with coordinates/major line width},
    xstep=\pgfkeysvalueof{/tikz/grid with coordinates/major step x},
    ystep=\pgfkeysvalueof{/tikz/grid with coordinates/major step y}
  },
  grid with coordinates/.cd,
  minor step x/.initial=.5,
  minor step y/.initial=.2,
  major step x/.initial=1,
  major step y/.initial=1,
  major line width/.initial=1pt,
}
\makeatother

\newlength\fheight
\newlength\fwidth
\usetikzlibrary{patterns}
\usetikzlibrary{plotmarks,patterns,decorations.pathreplacing,backgrounds,calc,arrows,arrows.meta,spy,matrix}

\pgfplotsset{compat=newest}

\pgfplotsset{mystyle/.style={%
        width=6cm,
        xmin=0,xmax=0.5,
        xtick={0,10,...,50}}}

\newcommand{\ceil}[1]{{\left\lceil #1\right\rceil}}

\newcommand{\argmax}[1]{\underset{#1}{\operatorname{arg}\,\operatorname{max}}\;}
\newcommand{\argmin}[1]{\underset{#1}{\operatorname{arg}\,\operatorname{min}}\;}

\newcommand{\figuresname}[1]{Figs.~}

\newcommand{\e}[1]{%
	\ifmmode\refstepcounter{equation}%
	  \eqno\mbox{\rm(\theequation)}\label{e:#1}%
	\else(\ref{e:#1})\fi}

\catcode`@=11
\newdimen\jot \jot=3pt
\def\openup{\afterassignment\@penup\dimen@=}
\def\@penup{\advance\lineskip\dimen@
  \advance\baselineskip\dimen@
  \advance\lineskiplimit\dimen@}
\def\eqalign#1{\null\,\vcenter{\openup\jot\m@th
  \ialign{\strut\hfil$\displaystyle{##}$&$\displaystyle{{}##}$\hfil
      \crcr#1\crcr}}\,}

\graphicspath{{./figure/}}

\def \fwidth{0.95\columnwidth}
\def \fheight {0.6\columnwidth}

\definecolor{color1}{HTML}{FFB14E}
\definecolor{color2}{HTML}{FA8775}
\definecolor{color3}{HTML}{EA5F94}
\definecolor{color4}{HTML}{CD34B5}
\definecolor{color5}{HTML}{9D02D7}
\definecolor{color6}{HTML}{0000FF}

\usepackage[acronyms,nonumberlist,nopostdot,nomain,nogroupskip]{glossaries}

\newacronym{3gpp}{3GPP}{3rd Generation Partnership Project}
\newacronym{adc}{ADC}{Analog to Digital Converter}
\newacronym{5g}{5G}{5th generation}
\newacronym{6g}{6G}{6th generation}
\newacronym{aimd}{AIMD}{Additive Increase Multiplicative Decrease}
\newacronym{am}{AM}{Acknowledged Mode}
\newacronym{amc}{AMC}{Adaptive Modulation and Coding}
\newacronym{aqm}{AQM}{Active Queue Management}
\newacronym{awgn}{AGWN}{Additive White Gaussian Noise}
\newacronym{balia}{BALIA}{Balanced Link Adaptation}
\newacronym{bdp}{BDP}{Bandwidth-Delay Product}
\newacronym{bf}{BF}{beamforming}
\newacronym{cc}{CC}{Congestion Control}
\newacronym{cav}{CAV}{Connected and Autonomous Vehicle}
\newacronym{cdf}{CDF}{Cumulative Distribution Function}
\newacronym{cn}{CN}{Core Network}
\newacronym{lbt}{LBT}{Listen Before Talk}
\newacronym{cqi}{CQI}{Channel Quality Information}
\newacronym{cp}{CP}{Control Plane}
\newacronym{csi}{CSI}{Channel State Information}
\newacronym{csirs}{CSI-RS}{Channel State Information - Reference Signal}
\newacronym{dc}{DC}{Dual Connectivity}
\newacronym{rb}{RB}{Resource Block}
\newacronym{dce}{DCE}{Direct Code Execution}
\newacronym{dci}{DCI}{Downlink Control Information}
\newacronym{udp}{UDP}{User Datagram Protocol}
\newacronym{dl}{DL}{downlink}
\newacronym{fcfs}{FCFS}{first-come-first-served}
\newacronym{dmr}{DMR}{Deadline Miss Ratio}
\newacronym{fspl}{FSPL}{free-space path loss}
\newacronym{dmrs}{DMRS}{DeModulation Reference Signal}
\newacronym{e2e}{E2E}{End-to-End}
\newacronym{ppp}{PPP}{Poission Point Process}
\newacronym{aoi}{AoI}{Area of Interest}
\newacronym{cpu}{CPU}{Central Processing Unit}
 \newacronym{gpu}{GPU}{Graphics Processing Unit}
 \newacronym{tpu}{TPU}{Tensor Processing Unit}
\newacronym{si}{SI}{Study Item}
\newacronym{ecn}{ECN}{Explicit Congestion Notification}
\newacronym{edf}{EDF}{Earliest Deadline First}
\newacronym{enb}{eNB}{eNodeB}
\newacronym{epc}{EPC}{Evolved Packet Core}
\newacronym{es}{ES}{Edge Server}
\newacronym{fdma}{FDMA}{Frequency Division Multiple Access}
\newacronym{fdd}{FDD}{Frequency Division Duplexing}
\newacronym{upa}{UPA}{Uniform Planar Array}
\newacronym[firstplural=Radio Access Technologies (RATs)]{rat}{RAT}{Radio Access Technology}
\newacronym{rt}{RT}{Ray Tracer}
\newacronym{fs}{FS}{Fast Switching}
\newacronym{isd}{ISD}{inter-site distance}
\newacronym{ftp}{FTP}{File Transfer Protocol}
\newacronym{gnb}{gNB}{Next Generation Node B}
\newacronym{harq}{HARQ}{Hybrid Automatic Repeat reQuest}
\newacronym{hetnet}{HetNet}{Heterogeneous Network}
\newacronym{hh}{HH}{Hard Handover}
\newacronym{hol}{HOL}{Head-of-Line}
\newacronym{ia}{IA}{Initial Access}
\newacronym{imt}{IMT}{International Mobile Telecommunication}
\newacronym{its}{ITS}{Intelligent Transport System}
\newacronym{iot}{IoT}{Internet of Things}
\newacronym{los}{LOS}{Line of Sight}
\newacronym{lte}{LTE}{Long Term Evolution}
\newacronym{m2m}{M2M}{Machine to Machine}
\newacronym{mac}{MAC}{Medium Access Control}
\newacronym{mc}{MC}{Multi-Connectivity}
\newacronym{mcs}{MCS}{Modulation and Coding Scheme}
\newacronym{mec}{MEC}{Mobile Edge Cloud}
\newacronym{mi}{MI}{Mutual Information}
\newacronym{mimo}{MIMO}{Multiple Input Multiple Output}
\newacronym{mmwave}{mmWave}{millimeter wave}
\newacronym{mptcp}{MPTCP}{Multipath TCP}
\newacronym{mr}{MR}{Maximum Rate}
\newacronym{mss}{MSS}{Maximum Segment Size}
\newacronym{mtd}{MTD}{Machine-Type Device}
\newacronym{mtu}{MTU}{Maximum Transmission Unit}
\newacronym{nfv}{NFV}{Network Function Virtualization}
\newacronym{vnf}{VNF}{Virtualization Network Function}
\newacronym{gv}{GV}{ground vehicle}
\newacronym{vec}{VEC}{Vehicular Edge Computing}
\newacronym{sdn}{SDN}{Software Defined Networking}
\newacronym{nlos}{NLOS}{Non Line of Sight}
\newacronym{nlosb}{NLOSb}{Building Non Line of Sight}
\newacronym{nlosv}{NLOSv}{Vehicle Non Line of Sight}
\newacronym{nr}{NR}{New Radio}
\newacronym{ofdm}{OFDM}{Orthogonal Frequency Division Multiplexing}
\newacronym{pdb}{PDB}{Packet Delay Budget}
\newacronym{pdcch}{PDCCH}{Physical Downlink Control Channel}
\newacronym{pdcp}{PDCP}{Packet Data Convergence Protocol}
\newacronym{pdsch}{PDSCH}{Physical Downlink Shared Channel}
\newacronym{pdu}{PDU}{Packet Data Unit}
\newacronym{pf}{PF}{Proportional Fair}
\newacronym{pgw}{PGW}{Packet Gateway}
\newacronym{phy}{PHY}{Physical}
\newacronym{prfs}{PRFS}{Physical Resource Frame Structure}
\newacronym{pbch}{PBCH}{Physical Broadcast Channel}
\newacronym{pscch}{PSCCH}{Physical Sidelink Control Channel}
\newacronym{pssch}{PSSCH}{Physical Sidelink Shared Channel}
\newacronym[plural=\gls{mme}s,firstplural=Mobility Management Entities (MMEs)]{mme}{MME}{Mobility Management Entity}
\newacronym{prb}{PRB}{Physical Resource Block}
\newacronym{pss}{PSS}{Primary Synchronization Signal}
\newacronym{pucch}{PUCCH}{Physical Uplink Control Channel}
\newacronym{pusch}{PUSCH}{Physical Uplink Shared Channel}
\newacronym{rach}{RACH}{Random Access Channel}
\newacronym{ran}{RAN}{Radio Access Network}
\newacronym{red}{RED}{Random Early Detection}
\newacronym{rf}{RF}{Radio Frequency}
\newacronym{rlc}{RLC}{Radio Link Control}
\newacronym{rlf}{RLF}{Radio Link Failure}
\newacronym{rrc}{RRC}{Radio Resource Control}
\newacronym{rrm}{RRM}{Radio Resource Management}
\newacronym{rr}{RR}{Round Robin}
\newacronym{rs}{RS}{Remote Server}
\newacronym{rsrp}{RSRP}{Reference Signal Received Power}
\newacronym{rss}{RSS}{Received Signal Strength}
\newacronym{rtt}{RTT}{Round Trip Time}
\newacronym{rw}{RW}{Receive Window}
\newacronym{rx}{RX}{Receiver}
\newacronym{sa}{SA}{standalone}
\newacronym{sack}{SACK}{Selective Acknowledgment}
\newacronym{sap}{SAP}{Service Access Point}
\newacronym{sch}{SCH}{Secondary Cell Handover}
\newacronym{scoot}{SCOOT}{Split Cycle Offset Optimization Technique}
\newacronym{sdma}{SDMA}{Spatial Division Multiple Access}
\newacronym{sinr}{SINR}{Signal to Interference plus Noise Ratio}
\newacronym{sm}{SM}{Saturation Mode}
\newacronym{snr}{SNR}{Signal to Noise Ratio}
\newacronym{son}{SON}{Self-Organizing Network}
\newacronym{ss}{SS}{Synchronization Signal}
\newacronym{srs}{SRS}{Sounding Reference Signal}
\newacronym{sss}{SSS}{Secondary Synchronization Signal}
\newacronym{tb}{TB}{Transport Block}
\newacronym{tcp}{TCP}{Transmission Control Protocol}
\newacronym{tdd}{TDD}{Time Division Duplexing}
\newacronym{tdma}{TDMA}{Time Division Multiple Access}
\newacronym{tfl}{TfL}{Transport for London}
\newacronym{tm}{TM}{Transparent Mode}
\newacronym{prr}{PRR}{Packet Reception Ratio}
\newacronym{trp}{TRP}{Transmitter Receiver Pair}
\newacronym{tti}{TTI}{Transmission Time Interval}
\newacronym{ttt}{TTT}{Time-to-Trigger}
\newacronym{tx}{TX}{Transmitter}
\newacronym{ue}{UE}{User Equipment}
\newacronym{ul}{UL}{uplink}
\newacronym{uml}{UML}{Unified Modeling Language}
\newacronym{um}{UM}{Unacknowledged Mode}
\newacronym{utc}{UTC}{Urban Traffic Control}
\newacronym{vm}{VM}{Virtual Machine}
\newacronym{rsrq}{RSRQ}{Reference Signal Received Quality}
\newacronym{rssi}{RSSI}{Received Signal Strength Indicator}
\newacronym{crs}{CRS}{Cell Reference Signal}
\newacronym{v2v}{V2V}{Vehicle-to-Vehicle}
\newacronym{v2i}{V2I}{Vehicle-to-Infrastructure}
\newacronym{v2n}{V2N}{Vehicle-to-Network}
\newacronym{v2x}{V2X}{Vehicle-to-Everything}
\newacronym{vn}{VN}{Vehicular Node}
\newacronym{dsrc}{DSRC}{Dedicated Short Range Communication}
\newacronym{ci}{CI}{context information}
\newacronym{voi}{VoI}{value of information}
\newacronym{gps}{GPS}{Global Positioning System}
\newacronym{qos}{QoS}{Quality of Service}
\newacronym{qoe}{QoE}{Quality of Experience}
\newacronym{ml}{ML}{Machine Learning}
\newacronym{ahp}{AHP}{Analytic Hierarchy Process}
\newacronym{lidar}{LIDAR}{Light Detection and Ranging}
\newacronym{sumo}{SUMO}{Simulation of Urban MObility}
\newacronym{wave}{WAVE}{Wireless Access in Vehicular Environment}
\newacronym{c-its}{C-ITS}{Connected Intelligent Transportation System}
\newacronym{dash}{DASH}{Dynamic Adaptive Streaming over HTTP}
\newacronym{http}{HTTP}{HyperText Transfer Protocol}
\newacronym{nt}{NT}{Non-Terrestrial}
\newacronym{ntc}{NTC}{non-terrestrial communication}
\newacronym{ntn}{NTN}{Non-Terrestrial Network}
\newacronym{haps}{HAPS}{High Altitude Platform Station}
\newacronym{hap}{HAP}{High Altitude Platform}
\newacronym{leo}{LEO}{Low Earth Orbit}
\newacronym{meo}{MEO}{Medium Earth Orbit}
\newacronym{geo}{GEO}{Geostationary Earth Orbit}
\newacronym{uav}{UAV}{Unmanned Aerial Vehicle}
\newacronym{nsat}{nSAT}{Nanosatellite}
\newacronym{ehf}{EHF}{extremely high-frequency}
\newacronym{ioe}{IoE}{Internet of Everyone}
\newacronym{gan}{GaN}{Gallium Nitride}
\newacronym{sci}{SCI}{Sidelink Control Information}
\newacronym{fr2}{FR2}{Frequency Range 2}
\newacronym{ula}{ULA}{Uniform Linear Array}
\newacronym{re}{RE}{Resource Element}
\newacronym{sl}{SL}{Sidelink}
\newacronym{sps}{SPS}{Semi Persistent Scheduling}
\newacronym{rc}{RC}{Reselection Counter}
\newacronym{rri}{RRI}{Resource Reservation Interval}
\newacronym{dbra}{DBRA}{directional beamformed resource allocation}

\newcommand\copyrightnotice{%
\begin{tikzpicture}[remember picture,overlay]
\node[anchor=south,yshift=15pt] at (current page.south) {\fbox{\parbox{\dimexpr\textwidth-\fboxsep-\fboxrule\relax}{
\footnotesize \textcopyright 2025 IEEE. Personal use of this material is permitted.
Permission from IEEE must be obtained for all other uses, in any current or future media,
including reprinting/republishing this material for advertising or promotional purposes,
creating new collective works, for resale or redistribution to servers or lists,
or reuse of any copyrighted component of this work in other works.}}};
\end{tikzpicture}
}

\begin{document}

\bstctlcite{IEEEexample:BSTcontrol}

\title{Sensing-Based Beamformed Resource Allocation in Standalone Millimeter-Wave Vehicular Networks\\
}

\author{\IEEEauthorblockN{Alessandro Traspadini,\textsuperscript{*} Anay Ajit Deshpande,\textsuperscript{*} Marco Giordani,\textsuperscript{*}}
\IEEEauthorblockN{Chinmay Mahabal,\textsuperscript{+} Takayuki Shimizu,\textsuperscript{+} Michele Zorzi\textsuperscript{*}}
\IEEEauthorblockA{\textsuperscript{*}Department of Information Engineering, University of Padova, Italy\\
\textsuperscript{+}R\&D InfoTech Labs,
Toyota Motor North America Inc., USA\\
Email: \texttt{\{traspadini,deshpande,giordani,zorzi\}@dei.unipd.it}} \texttt{\{chinmay.mahabal,takayuki.shimizu\}@toyota.com}
\thanks{This work was partially supported by the European Union under the Italian National Recovery and Resilience Plan (NRRP) Mission 4, Component 2, Investment 1.3, CUP C93C22005250001, partnership on “Telecommunications of the Future” (PE00000001 - program “RESTART”).}}

\maketitle

\copyrightnotice

\begin{abstract}
In 3GPP \gls{nr} \gls{v2x}, the new standard for next-generation vehicular networks, vehicles can autonomously select sidelink resources for data transmission, which permits network operations without cellular coverage. However, standalone resource allocation is uncoordinated, and is complicated by the high mobility of the nodes that may introduce unforeseen channel collisions (e.g., when a transmitting vehicle changes path) or free up resources (e.g., when a vehicle moves outside of the communication area). Moreover, unscheduled resource allocation is prone to the hidden node and exposed node problems, which are particularly critical considering directional transmissions.
In this paper, we implement and demonstrate a new channel access scheme for \gls{nr} \gls{v2x} in \gls{fr2}, i.e., at \gls{mmwave} frequencies, based on directional and beamformed transmissions along with \gls{sci} to select resources for transmission.
We prove via simulation that this approach can reduce the probability of collision for resource allocation, compared to a baseline solution that does not configure SCI transmissions.
\end{abstract}


\begin{tikzpicture}[remember picture,overlay]
\node[anchor=north,yshift=-10pt] at (current page.north) {\parbox{\dimexpr\textwidth-\fboxsep-\fboxrule\relax}{
\centering\footnotesize 
This paper has been accepted for publication in the 2025 IEEE International Conference on Communications (ICC). \textcopyright 2025 IEEE.\\
Please cite it as: A. Traspadini, A. A. Deshpande, M. Giordani, C. Mahabal, T. Shimizu, and M. Zorzi, "Sensing-Based Beamformed Resource Allocation in
Standalone Millimeter-Wave Vehicular Networks," in Proc. IEEE International Conference on Communications (ICC), 2025.}};
\end{tikzpicture}

\glsresetall
\section{Introduction}
In recent years, with the increasing number of vehicles on the road, the research community has been focusing on \glspl{cav} to improve safety and driving efficiency~\cite{gohar2021role, kombate2016internet}. 
The potential of \glspl{cav} can be fully unleashed through \gls{v2x} communication, which includes, for example, communication to and from cellular base stations (i.e., \gls{v2i}) and among vehicles (i.e., \gls{v2v}). 
On this front, the \gls{3gpp} is promoting the \gls{nr} \gls{v2x} standard to support \gls{v2x} communication~\cite{Zugno20Toward}. 
Along the lines of Rel-17 5G NR Uu specifications at 60 GHz, Rel-16 \gls{nr} \gls{v2x} involves sidelink transmissions in \gls{fr2}~\cite{lien20203gpp}, i.e., in the lower part of the \gls{mmwave} spectrum~\cite{matrouk2023energy, rasheed2020intelligent}. In particular, communication in the unlicensed bands in \gls{fr2}, especially at 60 GHz, is gaining momentum as an approach to alleviate spectrum scarcity, but also to promote better return of investment for service delivery. 

As far as resource allocation is concerned, \gls{nr} \gls{v2x} defines two modes for the selection of sidelink resources~\cite{garcia21tutorial}. In particular, in \gls{nr} \gls{v2x} Mode 2 vehicles can autonomously select their sidelink resources for data transmission, which permits network operations without cellular coverage. 
Specifically, vehicles determine the candidate set of resources based on \gls{sci} received from other vehicles, and on the \gls{rsrp} measured over the demodulation reference signals associated with the \gls{pssch} or the \gls{pscch}~\cite{38215}.
However, \gls{nr} \gls{v2x} in FR2 introduces additional challenges for resource allocation~\cite{tarafder2022mac}.
Most importantly, the severe path loss in these bands, and especially at 60 GHz due to oxygen absorption, require the vehicles to maintain directional transmissions via beamforming~\cite{giordani2017millimeter}. While this approach can reduce interference and collision, nodes may be listening to one specific direction, and receive partial \gls{sci} (or no \gls{sci} at all) in the rest of the angular space.
which complicates resource allocation. Moreover, channel dynamics in FR2 may cause \gls{sci} to become outdated.
Furthermore, \gls{nr} \gls{v2x} channel access is prone to the hidden node and exposed node problems.
On the one hand, communication in FR2 helps mitigate the exposed node problem, as a receiving node will detect transmissions only within its (narrow) beam.
On the other hand, the use of directional transmissions in FR2 exacerbates the hidden node problem, as vehicles may fail to detect transmissions outside of their beams.

Recently, the use of FR2 in V2X has been explored in the literature, mainly to support high-capacity transmissions for cooperative perception~\cite{bahbahani2022directional, su2022content,higuchi2019value}. However, most of these works do not consider the impact of directionality on the control plane, especially for resource allocation, or even exploit spatial diversity in FR2.


In this work, we implement and evaluate a resource allocation scheme for \gls{nr} \gls{v2x} Mode 2 at 60 GHz based on directional and beamformed communication. 
Specifically, each node receives SCI in the direction of transmission and in the paired direction, which indicates the resources being used for concurrent transmissions. These resources are labeled as blocked and will not be selected, thereby reducing the probability of collision.
While calibration results from the 3GPP only look at omnidirectional sensing, the novelty of this paper lies in the fact that we quantify resource allocation considering directional SCI transmissions, which is in line with the standardization activities on NR V2X towards Rel-19.
We run simulations as a function of the application rate, the density of vehicles, the antenna architecture, and the frame structure. We demonstrate that directional resource allocation can decrease by up to 75\% the probability of collision compared to some baseline solutions that do not use SCI and/or configure transmissions via large beams.

The rest of the paper is organized as follows.
\cref{sec:5gnr} presents the system model.
\cref{sec:resource_allocation} describes the proposed resource allocation scheme for \gls{nr} \gls{v2x} in FR2.
\cref{sec:simulation_results} presents our simulation results. \cref{sec:conclusion} concludes the paper with suggestions for future work.

\section{System model}\label{sec:5gnr}
In this section, we describe the \gls{nr} \gls{v2x} channel (\cref{sec:Phy}) and transmission (\cref{sub:transport}) models.

\begin{figure}[t!]
   \centering \includegraphics[width=0.4\textwidth]{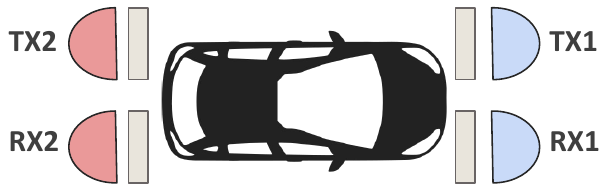}
   \caption{Vehicles antenna arrays setting.}
   \label{fig:vehicle}
\end{figure}

\subsection{Channel Model}
\label{sec:Phy}
Our scenario consists of a set of $K$ vehicles moving at a constant speed in an urban environment. 
Each vehicle is equipped with two transmitting antennas (TX1 and TX2) and two receiving antennas (RX1 and RX2), where TX1 and RX1 point forward, and TX2 and RX2 point backward, as illustrated in~\cref{fig:vehicle}.
Each transmit (receive) antenna is a \gls{ula} with $n_{tx}$ ($n_{rx}$) elements.

Based on the channel matrix  $\mathbf{H}_{i,j}\in \mathbb{C}^{n_{tx}} \times \mathbb{C}^{n_{rx}}$, which represents the channel between  transmitter $i$ and receiver $j$, we can derive the optimal beamforming vector $\mathbf{w}_{i,j}^{*}$ at the transmitter that maximizes the received \gls{sinr}.
This vector is selected within a set of codewords (i.e., the codebook) that depends on $n_{tx}$, i.e.,
\begin{equation}
\mathbf{w}_{i,j}^{*} = \argmax{\mathbf{w} \in C_{tx}} \mathbf{H}_{i,j} \mathbf{w},
\end{equation}
where $C_{tx}$ denotes the transmitter codebook.
On the other hand, we assume that the receiver points its beam in the boresight direction using beamforming vector $\mathbf{u}_{0}$.

As described in \cref{sub:res-all-nrv2x}, vehicles can sense the channel and receive SCI to reduce the probability of collision by avoiding selecting resources already reserved by other transmitters.
Then, vehicle $i$ receives \gls{sci}  using codeword $\mathbf{u}_{i,j}^s$,~i.e.,
\begin{equation}
    \mathbf{u}_{i,j}^s = \argmin{\mathbf{u} \in C_{rx}} ||\Bar{\Theta}_{\mathbf{u}} - \Bar{\Theta}_{\mathbf{w}_{i,j}^{*}} ||,
\end{equation}
where $\Bar{\Theta}_{\mathbf{u}} = [ \theta_{\mathbf{u}}, \phi_{\mathbf{u}} ]$ denotes the steering directions on the azimuth plane ($\theta_{\mathbf{u}}$) and on the elevation plane ($\phi_{\mathbf{u}}$) of the beamforming vector $\mathbf{u}$.
Thus, if the transmitter codebook and the receiver codebook are the same (i.e., $C_{rx}=C_{tx}$), vehicle~$i$ selects the same codeword for both data transmission and sensing (i.e., $\mathbf{u}_{i,j}^s = \mathbf{w}_{i,j}^{*}$).

We can evaluate the probability of collision as the probability that two or more pairs of nodes select the same resources for transmission.
For each transmission from vehicle $i$ to vehicle $j$, we define the set $C_{(i,j)}$ of transmitter-receiver pairs $(l,m)$ that share the same resources as $(i,j)$.
The \gls{sinr} between transmitter $i$ and receiver $j$ at distance $d$  is given~by
\begin{equation}
\Gamma_{i,j}^{d} = \frac{\mathbf{u}_{0} \cdot \mathbf{H}_{i,j} \cdot \mathbf{w}_{i,j}^{*}}{ k \cdot T \cdot B + \sum_{(l,m)\in C_{(i,j)}} (\mathbf{u}_{0} \cdot \mathbf{H}_{l,j} \cdot \mathbf{w}_{l,m}^{*})},
\label{eq:sinr}
\end{equation}
where $k$ is the Boltzmann constant, $T$ is the noise temperature, $B$ is the channel bandwidth, $\mathbf{H}$ is the channel matrix, $\mathbf{w}_{i,j}^{*}$ and $\mathbf{w}_{l,m}^{*}$ are the optimal beamforming vectors of the transmitters (both intended and interfering, respectively), and $\mathbf{u}_0$ is the beamforming vector of the intended receiver.
The expression in~\cref{eq:sinr} accounts for the interference due to concurrent transmissions from vehicle $l$ to vehicle $m$, which overlap with the intended transmission from $i$ to $j$.
Then, vehicles $i$ and $j$ are in coverage and can successfully receive and decode data transmissions if $\Gamma_{i,j}^{d} \geq \gamma_{th}$, where $\gamma_{th}$ is the \gls{sinr} threshold.

\subsection{Transmission Model}
\label{sub:transport}

The \gls{sci} is transmitted over the \gls{pscch}, and can be decoded by sensing the channel~\cite{lien20203gpp}. 
As described in~\cref{sub:res-all-nrv2x}, SCI is used to select resources, i.e., \glspl{prb}, to be used for data transmissions over the \gls{pssch}. From~\cite{38214}, a \gls{prb} consists of $N^{prb}_{sr}$ subcarriers in frequency and $N_{sy}$ \gls{ofdm} symbols in time, i.e., in a slot (excluding Automatic Gain Control (AGC) symbol and a guard symbol). Overall, a group of $N_{prb}^{sh}$ \glspl{prb} create a subchannel, and there are $N_{sh}$ consecutive subchannels in the available bandwidth.
Now, the number of \glspl{re} allocated for transmission per slot, i.e., excluding \glspl{prb} used for control, is:
\begin{equation}
    N_{re} = (N^{prb}_{sr} \cdot N_{sy} - N_{dmrs}) \cdot N_{prb}^{sh} \cdot N_{sh} - N^{prb}_{sr} \cdot N_{sci-1},
    \label{eq:N_re}
\end{equation}
where $N_{dmrs}$ is the number of \glspl{re} allocated to the \gls{dmrs} for channel estimation,  and $N_{sci-1}$ is the number of \glspl{prb} used for \gls{pscch}.
Based on~\cref{eq:N_re}, the number of bits per slot that can be allocated for transmission ($s_{s}$) is
\begin{equation}
    s_{s} = N_{re} \cdot R \cdot Q_m \cdot n_{l} - N_{sci-2},
    \label{eq:s_tb}
\end{equation}
where $R$ is the code rate, $Q_m$ is the modulation order, $n_{l}$ is the number of layers, and $N_{sci-2}$ is the overhead for the 2nd-stage \gls{sci}, which is transmitted over the \gls{pssch}~\cite{lien20203gpp}.

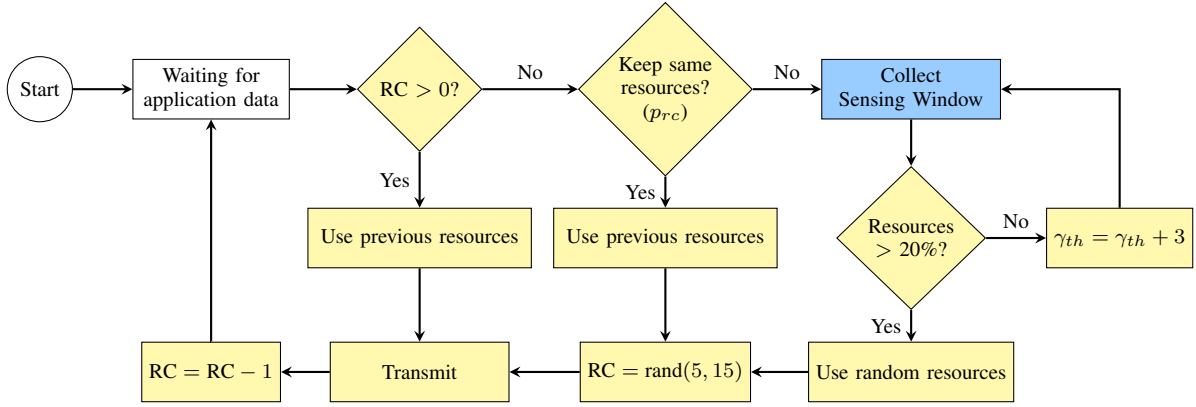
\begin{figure*}[t!]
    \centering
    \tikzstyle{startstop} = [circle, minimum size=0.8cm, draw=black, text centered, fill=yellow!40]
\tikzstyle{process} = [rectangle, minimum width=1cm, minimum height=0.8cm, inner sep=0.1cm, draw=black, text centered, fill=yellow!40]
\tikzstyle{decision} = [diamond, minimum width=1.2cm, minimum height=1.2cm, inner sep=0.1cm, draw=black, text centered, fill=yellow!40]
\tikzstyle{arrow} = [thick,->,>=stealth]

\definecolor{color_sensing}{RGB}{153, 204, 255}

\begin{tikzpicture}[node distance=0.8cm, font=\footnotesize, scale=0.99, transform shape]

    \node (Start) [startstop, fill= white] {Start};
    \node (AppData) [process, right of=Start, xshift=1.5cm, inner sep=0.05cm, fill=white] {\parbox{2cm}{\centering Waiting for \\ application data}};
    \node (RC) [decision, right of=AppData, xshift=2cm] {RC $> 0$?};
    \node (Pc) [decision, right of=RC, xshift=2.5cm, minimum width=1cm, minimum height=1cm, text width=1.3cm, inner sep=0.04cm, align=center] {Keep same resources? ($p_{rc}$)};
    
    \node (Gather) [process, right of=Pc, xshift=2.5cm, fill=color_sensing] {\parbox{2.2cm}{\centering Collect\\ Sensing Window}};

    \node (ResPrev1) [process, below of=RC, yshift=-1.2cm] {\parbox{2.8cm}{\centering Use previous resources}};
    \node (ResPrev2) [process, below of=Pc, yshift=-1.2cm] {\parbox{2.8cm}{\centering Use previous resources}};

    \node (NumRes) [decision, below of=Gather, yshift=-1.2cm, minimum width=2cm, minimum height=1cm, text width=1.2cm, inner sep=0.05cm, align=center] {Resources $>$ 20\%?};

    \node (Tx) [process, below of=ResPrev1, yshift=-1cm] {\parbox{2.2cm}{\centering Transmit}};

    \node (RandRC) [process, below of=ResPrev2, yshift=-1cm] {$\text{RC} = \text{rand}(5,15)$};
    
    \node (RandomRes) [process, right of=RandRC, xshift=2.5cm] {Use random resources};
    \node (UpdateRC) [process, below of=AppData, yshift=-3cm] {$\text{RC} = \text{RC} - 1$};
    
    \node (SINR) [process, right of=NumRes, xshift=2cm] {$\gamma_{th} = \gamma_{th} + 3$};
    
    \draw [arrow] (Start) -- (AppData);
    \draw [arrow] (AppData) -- (RC);
    \draw [arrow] (RC) -- node[left] {Yes} (ResPrev1);
    \draw [arrow] (RC.east) -- node[anchor=south] {No} (Pc.west);
    \draw [arrow] (Pc)  -- node[above] {No} (Gather);
    \draw [arrow] (Pc) -- node[left] {Yes} (ResPrev2);
    \draw [arrow] (Gather) -- (NumRes);
    \draw [arrow] (NumRes) -- node[left] {Yes} (RandomRes);
    \draw [arrow] (RandomRes.west) -- (RandRC.east);
    \draw [arrow] (ResPrev1) -- (Tx);
    \draw [arrow] (ResPrev2) -- (RandRC);
    \draw [arrow] (RandRC) -- (Tx.east);
    \draw [arrow] (NumRes) -- node[above] {No} (SINR.west);
    \draw [arrow] (SINR.north) |- (Gather.east);
    \draw [arrow] (Tx) -- (UpdateRC);
    \draw [arrow] (UpdateRC.north) -| (AppData.south);
\end{tikzpicture}
    \caption{Flowchart for the Selection Window in \acrshort{dbra}.}
    \label{fig:slots_flow}
\end{figure*}

\section{Resource Allocation in NR V2X}
\label{sec:resource_allocation}
In this section we describe resource allocation (\cref{sub:res-all-nrv2x}) and directional beamformed resource allocation (\cref{sub:res-beam-nrv2x}) in \gls{nr} \gls{v2x}.
\subsection{Resource Allocation in NR V2X Mode 2}
\label{sub:res-all-nrv2x}
The \gls{nr} \gls{v2x} standard defines two resource allocation modes (namely Mode~1 and Mode~2) for \gls{v2x} sidelink communication~\cite{3gppTR38885}.
In Mode~1 (similar to Mode~3 in \gls{lte} \gls{v2x}), the selection of the resources is coordinated by a \gls{gnb}, which reduces the probability of collision but requires vehicles to be in coverage of a cellular infrastructure.
In Mode~2 (similar to Mode~4 in \gls{lte} \gls{v2x}), vehicles can autonomously select their sidelink resources without the coordination of a \gls{gnb}, which permits network operations without coverage.
Mode~2 supports both dynamic and \gls{sps}, with the latter reserving resources for multiple consecutive transmissions.

In \gls{nr} \gls{v2x} Mode 2 \gls{sps}, each vehicle reserves resources
for data transmissions for a number of consecutive \glspl{tb} given by the \gls{rc}.
The \gls{rc} is based on the value of the \gls{rri}: if RRI $<100$ ms, the RC is randomly selected within the interval $[5C,15C]$, where $C=100/\max(20,\text{RRI})$; if RRI $\geq100$ ms, the RC is randomly selected within the interval $[5,15]$~\cite{lusvarghi2023comparative}.
We define two quantities:
\begin{itemize}
    \item The ``Sensing Window,'' i.e., the set of channel resources (in the time and frequency domains) that the vehicle ``senses,'' i.e., listens to. In this phase, each vehicle receives the 1st-stage \gls{sci}, transmitted by other vehicles in the Sensing Window resources.
    Specifically, it indicates the sidelink resources that other vehicles have reserved for their \gls{sci} and \gls{tb} transmissions in the \gls{pscch} and \gls{pssch}, respectively.
    \item The ``Selection Window,'' i.e., the set of candidate (not reserved) time and frequency resources that may be used for transmission. The Selection Window is identified based on the Sensing Window, which is derived from the decoded 1st-stage SCI and its \gls{rsrp}, measuring the quality of the signal.
\end{itemize}

\subsection{Directional Beamformed Resource Allocation}
\label{sub:res-beam-nrv2x}
\begin{figure}
    \centering
    \includegraphics[width=\linewidth]{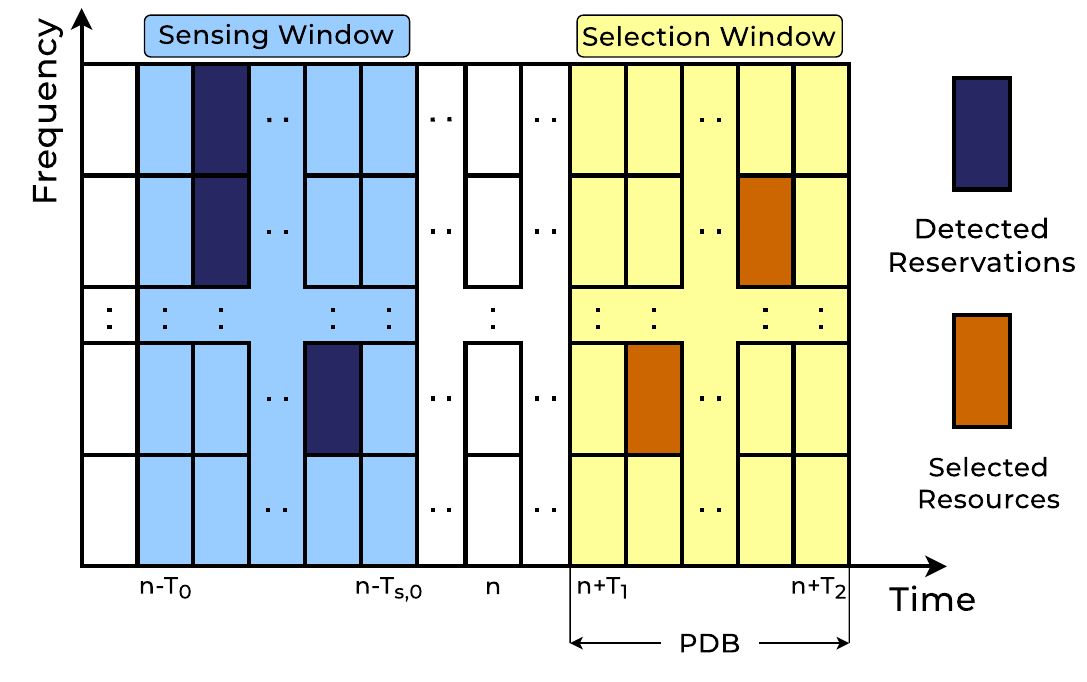}
    \captionsetup{skip=0.1pt}
    \caption{Illustration of the Sensing and Selection Windows in \acrshort{dbra}.\vspace{-0.5cm}}
    \label{fig:slots_window}
\end{figure}

Resource allocation in \gls{nr} \gls{v2x}  Mode 2 as explained in \cref{sub:res-all-nrv2x} does not consider directionality and spatial diversity while selecting resources, which is essential in \gls{fr2}. 
In particular, resource allocation is based on the assumption that vehicles can sense the channel via omnidirectional communication, which permits to receive SCI from most neighbors, thus reducing the probability of collision.
In FR2, instead, it may be essential to establish directional communication even for SCI transmissions, in order to compensate for the higher path loss at these frequencies, and extend the range via beamforming.
Directional communication, however, requires precise alignment of the transmitting and receiving beams, and may suffer from the hidden node problem since a vehicle can only receive \gls{sci} if it is within the direction of the transmitter.
To address these challenges, we define a \gls{dbra} scheme for NR V2X that takes into account the directionality and spatial diversity introduced by using antenna arrays for beamforming in FR2.
We define the following procedure, also illustrated in \cref{fig:slots_window,fig:slots_flow}.

\begin{itemize}
    \item Definition of the Sensing Window: At slot $n$, the Sensing Window is defined in the range $[n-T_0,n-T_{s,0}]$, where $T_0$ is a preconfigured integer, equivalent to $1100$ ms or $100$ ms in number of slots, and $T_{s,0}$ is the sensing processing time in number of slots~\cite{garcia21tutorial}.
    Ideally, exhaustive search should be used to receive the SCI through a predefined codebook of directions covering the entire angular space~\cite{giordani2018tutorial}.
    However, this approach would introduce a large delay required to sweep all possible directions during the sensing procedure. 
    Additionally, it could intensify the exposed node problem by detecting resources that would not cause interference.
    Therefore, we restrict SCI transmission to two specific directions, as opposed to omnidirectional sensing in baseline NR V2X Mode~2.
    These directions are the ``primary direction,'' determined based on the receiver's location, and the ``paired direction,'' which is the opposite of the primary.
The received SCI is used to identify the set of resources that cannot be selected for subsequent transmissions. 
Specifically, resources are marked as ``reserved'' if the
relative \gls{rsrp} is higher than a threshold, meaning that there might be other concurrent transmissions in those~resources.
\item Definition of the Selection Window: 
The Selection Window is defined in the range $[n+T_1,n+T_2]$, where $T_1$ is the number of slots required for selecting resources, and $T_2$ is less than or equal to the \gls{pdb} in slots to ensure timely transmission.
In fact, the \gls{pdb} is determined by the \gls{v2x} application, and refers to the latency deadline by which \glspl{tb} must be transmitted~\cite{molina2023impact}.
 The PDB is critical in NR V2X to ensure timely delivery of data, especially for safety-related applications.
As the \gls{pdb} increases, we can extend the length of the Selection Window, so the set of available resources for transmission.
Within the Selection Window, the vehicle randomly selects a set of $N_{s}=s_{d}/s_{s}$ free resources (slots) based on the Sensing Window, where $s_d$ is the size of the data to send, and $s_{s}$ is given in~\cref{eq:s_tb}.
 \item As described in \cref{sub:res-all-nrv2x}, resources are selected for a number of consecutive TBs given by the RC, which is specific to each receiver.
At each transmission, the RC is decreased by 1. When it reaches 0, the same set of resources is selected with probability $p_{rc}$, while with probability $1-p_{rc}$ a new Sensing Window is defined to select a new set of resources.
\end{itemize}

\begin{table}[t!]
\centering
\vspace{0.3cm}
\caption{Simulation parameters.}
\label{tab:parameters}
\begin{tabular}{@{}lcc@{}}
\toprule
\textbf{Parameter} & \textbf{Symbol} & \textbf{Value} \\ \midrule
OFDM symbols per slot & $N_{sy}$ & 12 \\
Subcarriers per PRB & $N^{prb}_{sr}$ & 12 \\
REs for DMRS & $N_{dmrs}$ & 18 \\
PRBs per subchannel & $N_{prb}^{sh}$ & 25 \\
Number of subchannels & $N_{sh}$ & \{1, 2, 3, 4\} \\
PRBs used for PSCCH & $N_{sci-1}$ & 50 \\
2nd-stage SCI overhead [Bits] & $N_{sci-2}$ & 48 \\
Code rate & $R$ & 0.7 \\
Modulation order & $Q_m$ & 6 \\
Number of layers & $n_l$ & 1 \\
Resource Reservation Interval [ms] & RRI & 100 \\ 
Application rate [Mbps] & $D$ & \{5.7, 20, 40\} \\ \midrule
TX antenna elements & $n_{tx}$ & \{4, 16, 64\} \\
RX antenna elements & $n_{rx}$ & 2 \\
Total bandwidth [MHz] & $B$ & 400 \\
Transmit power [dBm] & $P_{tx}$ & 23 \\
Noise temperature [K] & $T$ & 300 \\
SNR threshold [dB] & $\gamma_{th}$ & 0 \\
Resource reuse probability & $p_{rc}$ & 0 \\
Available resources [\%] & $X$ & 20 \\
Carrier frequency [GHz] & $f_c$ & 60 \\\bottomrule
\end{tabular}
\end{table}

\begin{figure*}[t!]
    \centering
    \subfloat[1W-LD]{
        \includegraphics[width=0.2\textwidth]{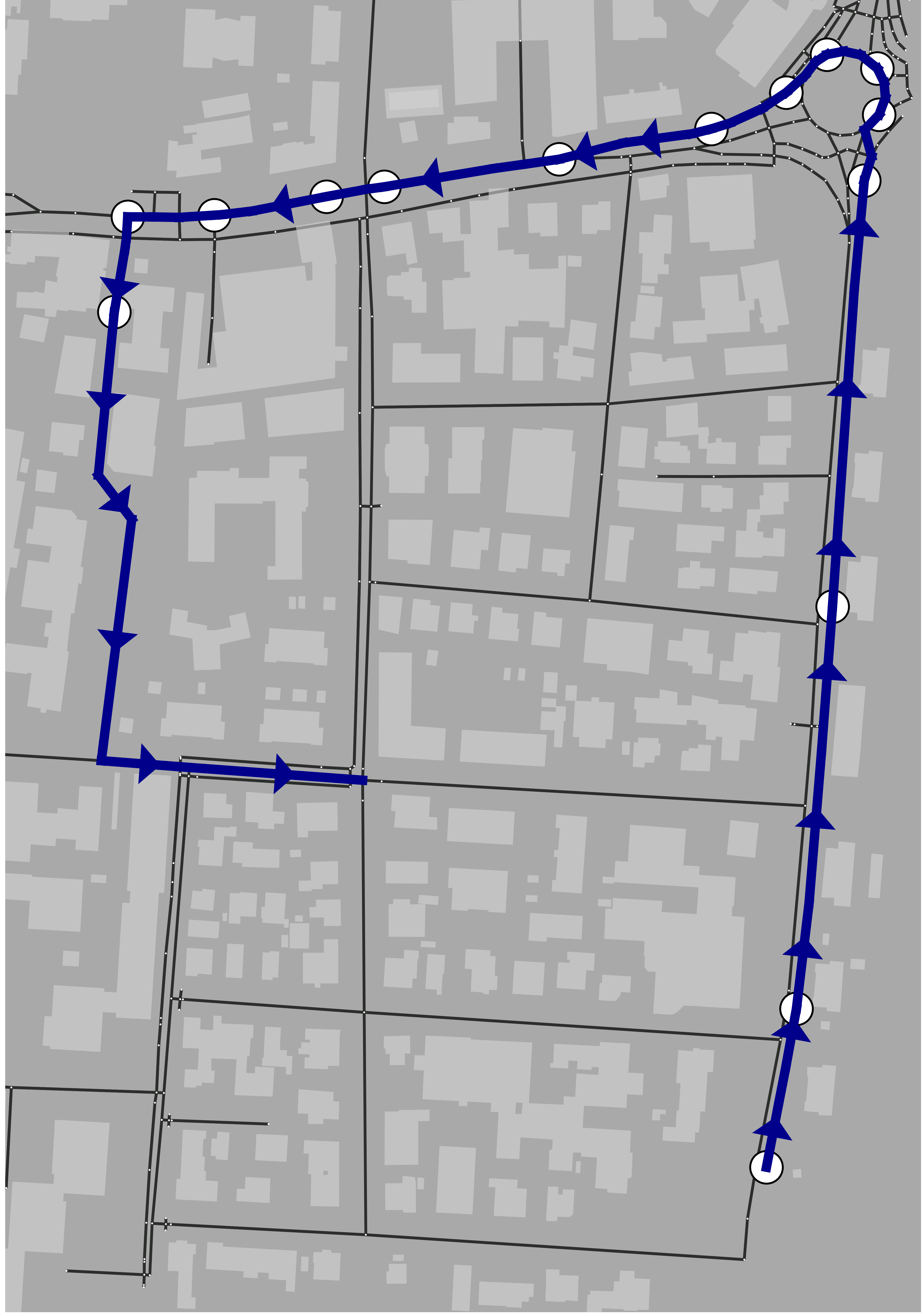}
        \label{fig:map_1w-ld}
    }\hfill
    \subfloat[1W-HD]{
        \includegraphics[width=0.2\textwidth]{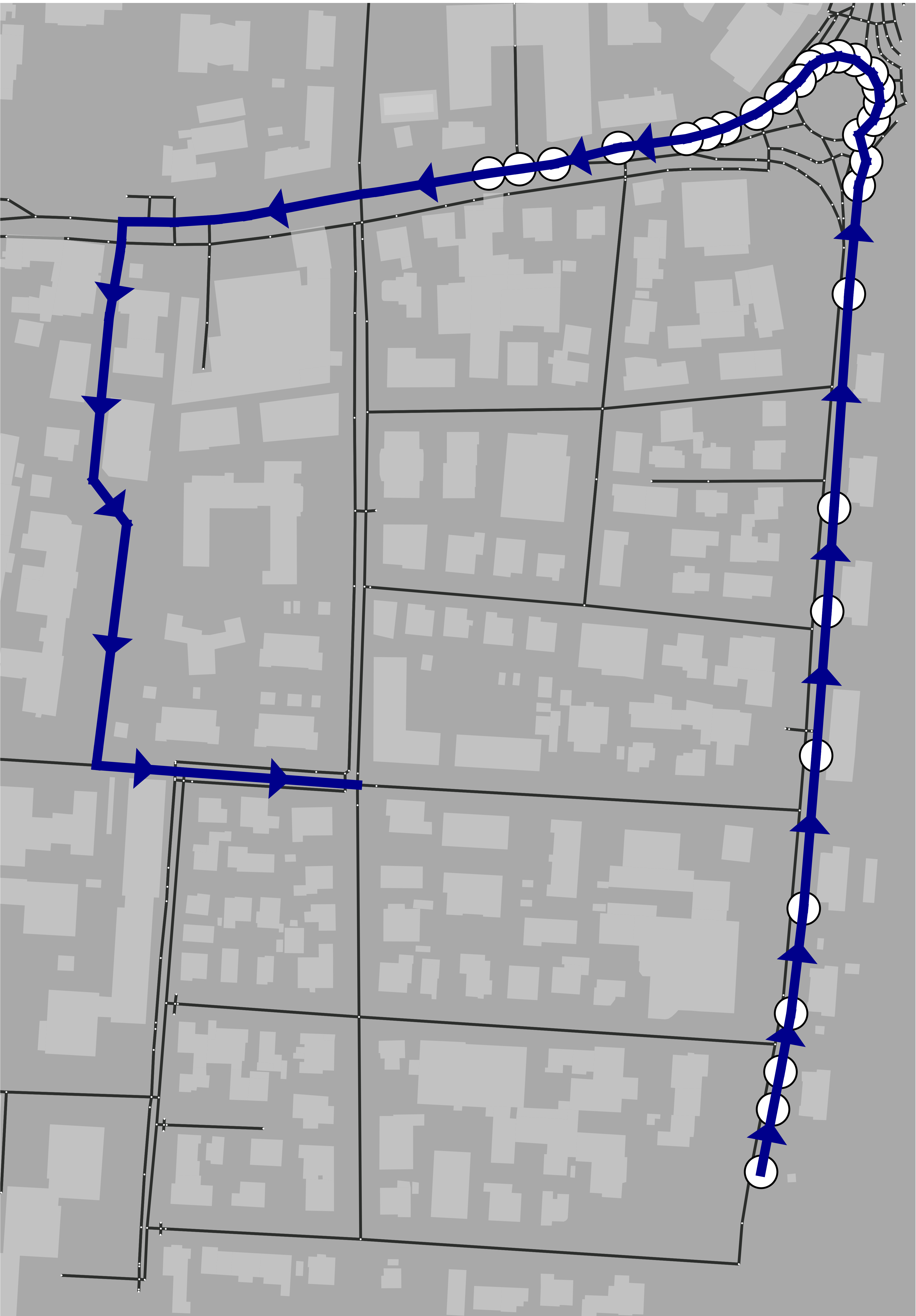}
        \label{fig:map_1w-hd}
    }\hfill
    \subfloat[2W-LD]{
       \includegraphics[width=0.2\textwidth]{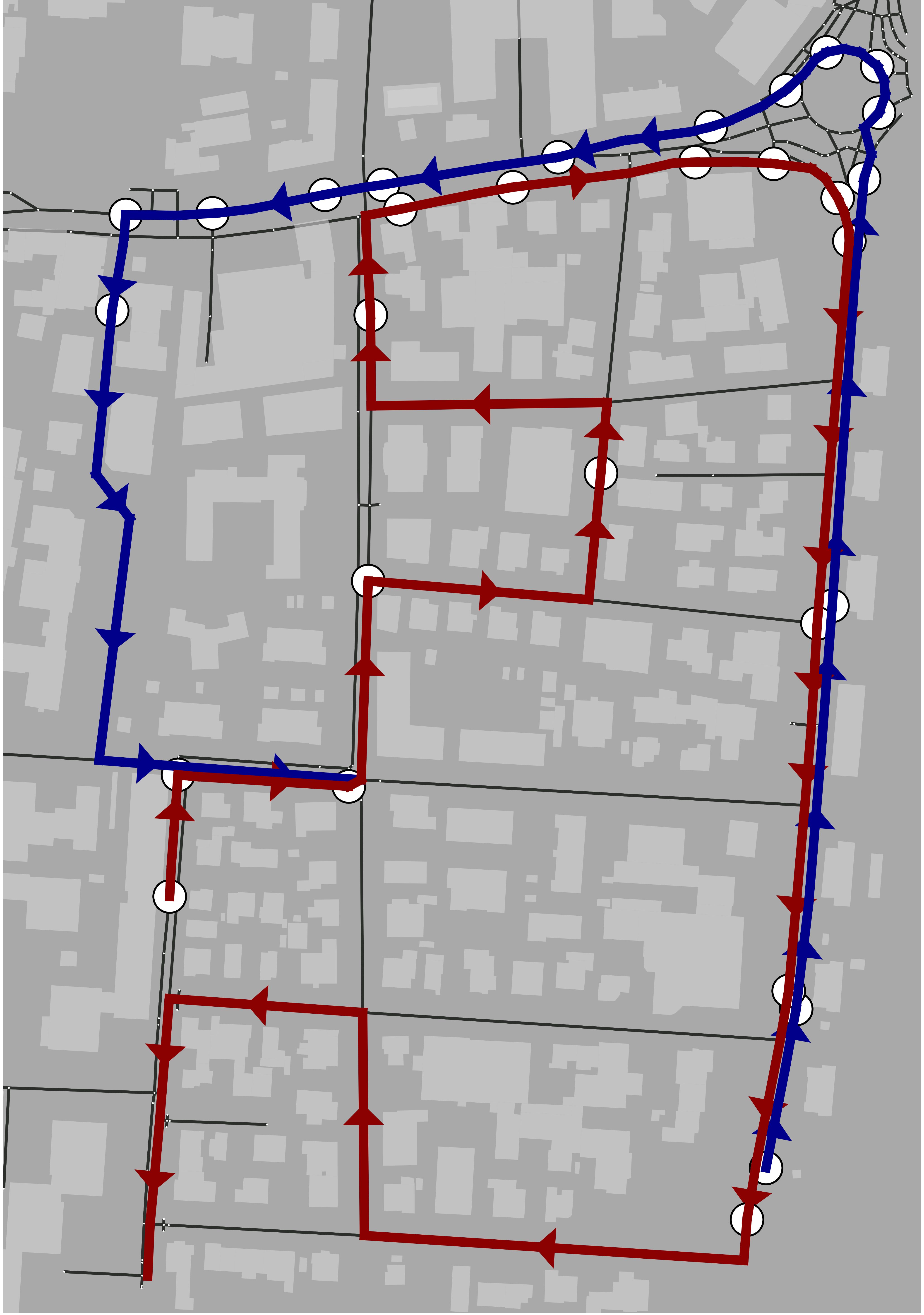}
       \label{fig:map_2w-ld}
    }\hfill
    \subfloat[2W-HD]{
       \includegraphics[width=0.2\textwidth]{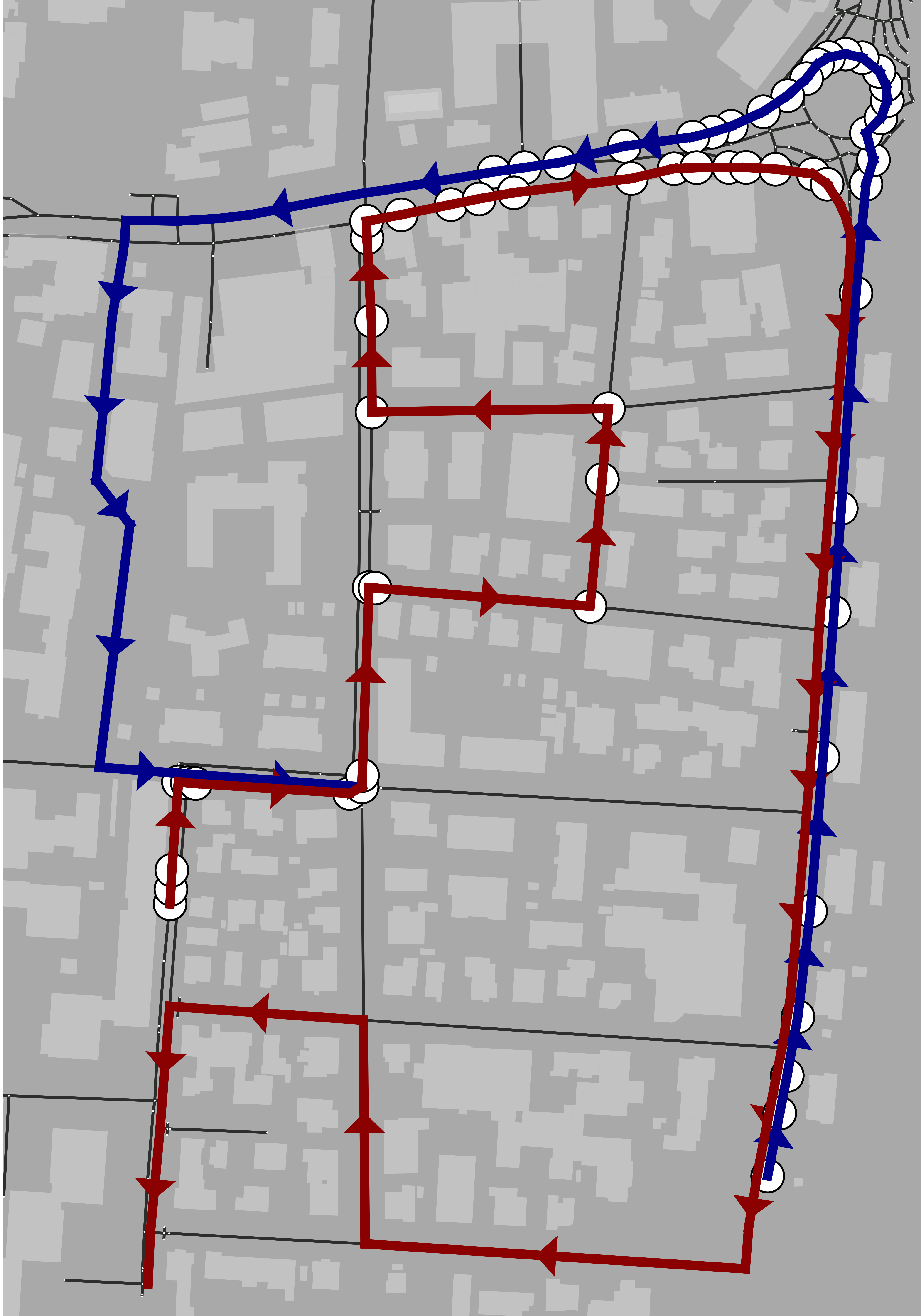}
       \label{fig:map_2w-hd}
    }
    \caption{Maps of the urban scenarios. White dots indicate the initial positions of the vehicles, while red and blue lines represent the routes of the vehicles in the corresponding lanes.}
    \label{fig:map}
\end{figure*}

\section{Performance Evaluation}
\label{sec:simulation_results}
In this section we present our simulation parameters (\cref{sec:simulation_scenario}) and numerical results (\cref{sub:results}).

\subsection{Simulation Parameters}
\label{sec:simulation_scenario}

The simulation scenario is implemented in Python, and the physical layer is designed based on \gls{3gpp} Rel-18 specifications for \gls{nr} \gls{v2x}.
The positions of the vehicles are obtained from OpenStreetMap data, specifically for a small area within the city of Padova, Italy.
Vehicles move along the grid, defined by OpenStreetMap, following a constant velocity model, traveling at a fixed speed of 10~m/s.
Each vehicle transmits data frames (e.g., generated from onboard sensors) of size $s_d\in\{0.57,2,4\}$ Mb at a rate of $r=10$ fps, to up to $K_{rx}$ neighboring vehicles (set to 5 in our simulations), therefore the resulting application rate is $D=s_d\cdot r \in\{5.7,20,40\}$ Mbps. 

We consider \gls{v2v} communication and \gls{sci} transmissions in FR2 in the unlicensed bands at 60~GHz, with a total bandwidth of $B=400$ MHz. 
The transmit (receive) antenna array is a \gls{ula} with $n_{tx}\in \{4, 16, 64\}$ ($n_{rx}=2$) elements. The radiation pattern of both is based on the \gls{3gpp} Technical Report in~\cite{38901}.

The rest of the simulation parameters are reported in~\cref{tab:parameters}. 

Our simulator implements and uses the NVIDIA Sionna \gls{rt} to compute the channel and characterize the propagation of the signal~\cite{sionna}. This \gls{rt} is \gls{gpu} accelerated, which improves the performance in terms of speed and scalability.
For each pair of vehicles, we compute the channel matrix $\mathbf{H}$ using Sionna, and derive the optimal beamforming vector $\mathbf{w}^{*}$ to maximize the \gls{sinr}, as described in~\cref{sec:Phy}.

We analyze four different scenarios in an urban environment, as illustrated in \cref{fig:map}:

\begin{itemize}
    \item {{One-Way Low Density (1W-LD)}} in~\cref{fig:map_1w-ld} and {{One-Way High Density (1W-HD)}} in~\cref{fig:map_1w-hd}: These scenarios represent a one-way urban street with $K=15$ and $30$ vehicles, respectively.
    \item {{Two-Way Low Density (2W-LD)}} in~\cref{fig:map_2w-ld} and {{Two-Way High Density (2W-HD)}} in~\cref{fig:map_2w-hd}: These scenarios represent a two-way urban street with 15 and 30 vehicles traveling in each direction, therefore $K=30$ and $60$, respectively.
\end{itemize}

\cref{tab:vehicle_distances} reports the statistics (number of vehicles, and average and standard deviation of the distance between pairs of vehicles) for the considered urban scenarios.
As expected, as the number of vehicles increases, the average distance and variance decrease accordingly.

Additionally, we explored four different \gls{prfs} configurations:

\begin{itemize}
    \item {\gls{prfs}-1}: Numerology $3$ with $N_{sh}=4$.
    \item {\gls{prfs}-2}: Numerology $4$ with $N_{sh}=3$.
    \item {\gls{prfs}-3}: Numerology $5$ with $N_{sh}=2$.
    \item {\gls{prfs}-4}: Numerology $6$ with $N_{sh}=1$.
\end{itemize}


Notably, $N_{sh}$ is a function of the numerology.
As the numerology decreases, with fixed bandwidth, the slot duration increases (from 31.25 $\mu$s for numerology 6 to 125 $\mu$s for numerology 3~\cite{38214}), so the number of slots in the time domain decreases. 
 As a result, more subchannels are required in the frequency domain to compensate this condition, and maintain the same total number of resources (from $N_{sh}=1$ for numerology 6 to $N_{sh}=4$ for numerology~3).

We compare the following resource allocation schemes:

\begin{itemize}
\item {\gls{dbra}}: It is the directional beamformed resource allocation scheme presented in~\cref{sub:res-beam-nrv2x}, where \gls{sci} is transmitted in both the primary and paired directions.
\item {DBRA-O}: It is the same as \gls{dbra}, but \gls{sci} is transmitted only in the primary direction to reduce interference.
\item {RRA} (benchmark): Resources are selected randomly in the Selection Window.
\end{itemize}

Each scheme requires a specific number of resources to be allocated for \gls{sci} transmissions, as reported  in~\cref{tab:resource_allocation}.

\begin{figure}[t!]
        \begin{subfigure}[b]{\linewidth}
	   \centering
%
%

\definecolor{black25}{RGB}{25,25,25}
\definecolor{mediumaquamarine102194165}{RGB}{102,194,165}
\definecolor{limegreen3122331}{RGB}{31,223,31}
\definecolor{silver}{RGB}{192,192,192}
\definecolor{lightgreen178223138}{RGB}{178,223,138}
\definecolor{lightsteelblue173203219}{RGB}{173,203,219}
\definecolor{steelblue31120180}{RGB}{31,120,180}

\definecolor{color1}{RGB}{0, 76, 153}
\definecolor{color2}{RGB}{34, 139, 34}
\definecolor{color3}{RGB}{255, 140, 0}

\begin{tikzpicture}
\pgfplotsset{every tick label/.append style={font=\scriptsize}}

\pgfplotsset{compat=1.11,
	/pgfplots/ybar legend/.style={
		/pgfplots/legend image code/.code={%
			\draw[##1,/tikz/.cd,yshift=-0.25em]
			(0cm,0cm) rectangle (10pt,0.6em);},
	},
}

\begin{axis}[%
width=0,
height=0,
at={(0,0)},
scale only axis,
xmin=0,
xmax=0,
xtick={},
ymin=0,
ymax=0,
ytick={},
axis background/.style={fill=white},
legend style={legend cell align=left,
              align=center,
              draw=white!15!black,
              at={(0.5, 1.3)},
              anchor=center,
              /tikz/every even column/.append style={column sep=1em}},
legend columns=3,
]
\addplot[ybar,ybar legend,draw=black,fill=color1,line width=0.08pt]
table[row sep=crcr]{%
	0	0\\
};
\addlegendentry{DBRA}

\addplot[ybar legend,ybar,draw=black,fill=color2,line width=0.08pt]
  table[row sep=crcr]{%
	0	0\\
};
\addlegendentry{DBRA-O}

\addplot[ybar legend,ybar,draw=black,fill=color3,line width=0.08pt]
  table[row sep=crcr]{%
	0	0\\
};
\addlegendentry{RRA}

\end{axis}
\end{tikzpicture}%
	   \end{subfigure}
    \vskip 0.1cm
    \begin{subfigure}[b]{\linewidth}
        \centering
\begin{tikzpicture}

\definecolor{darkslategray38}{RGB}{38,38,38}
\definecolor{lightgray204}{RGB}{204,204,204}
\definecolor{color1}{RGB}{0, 76, 153}
\definecolor{color2}{RGB}{34, 139, 34}
\definecolor{color3}{RGB}{255, 140, 0}

\begin{axis}[
width = \textwidth,
height = 5cm,
axis line style={lightgray204},
tick align=outside,
x grid style={lightgray204},
xlabel=\textcolor{darkslategray38}{Number of antenna elements at the transmitter ($n_{tx}$)},
xmajorgrids,
ymajorgrids,
xmajorticks=true,
xmin=-0.5, xmax=2.5,
xtick style={color=darkslategray38},
xtick={0,1,2},
xticklabels={4, 16, 64},
y grid style={lightgray204},
ylabel=\textcolor{darkslategray38}{SINR ($\Gamma^d$) [dB]},
ymajorticks=true,
ymin=-50, ymax=85,
ytick style={color=darkslategray38},
ytick={-40,-20,0,20,40,60,80},
yticklabels={
  \(\displaystyle {\ensuremath{-}40}\),
  \(\displaystyle {\ensuremath{-}20}\),
  \(\displaystyle {0}\),
  \(\displaystyle {20}\),
  \(\displaystyle {40}\),
  \(\displaystyle {60}\),
  \(\displaystyle {80}\)
}
]
\path [draw=black, fill=color1, line width=0.48pt]
(axis cs:-0.397333333333333,11.7289755418051)
--(axis cs:-0.136,11.7289755418051)
--(axis cs:-0.136,35.2825253928013)
--(axis cs:-0.397333333333333,35.2825253928013)
--(axis cs:-0.397333333333333,11.7289755418051)
--cycle;
\path [draw=black, fill=color2, line width=0.48pt]
(axis cs:-0.130666666666667,9.11200304474079)
--(axis cs:0.130666666666667,9.11200304474079)
--(axis cs:0.130666666666667,34.3689905730001)
--(axis cs:-0.130666666666667,34.3689905730001)
--(axis cs:-0.130666666666667,9.11200304474079)
--cycle;
\path [draw=black, fill=color3, line width=0.48pt]
(axis cs:0.136,6.29217439327446)
--(axis cs:0.397333333333333,6.29217439327446)
--(axis cs:0.397333333333333,32.7440241294636)
--(axis cs:0.136,32.7440241294636)
--(axis cs:0.136,6.29217439327446)
--cycle;
\path [draw=black, fill=color1, line width=0.48pt]
(axis cs:0.602666666666667,16.060533220124)
--(axis cs:0.864,16.060533220124)
--(axis cs:0.864,41.0328867512672)
--(axis cs:0.602666666666667,41.0328867512672)
--(axis cs:0.602666666666667,16.060533220124)
--cycle;
\path [draw=black, fill=color2, line width=0.48pt]
(axis cs:0.869333333333333,12.9477614391104)
--(axis cs:1.13066666666667,12.9477614391104)
--(axis cs:1.13066666666667,40.1701089263648)
--(axis cs:0.869333333333333,40.1701089263648)
--(axis cs:0.869333333333333,12.9477614391104)
--cycle;
\path [draw=black, fill=color3, line width=0.48pt]
(axis cs:1.136,9.77542317560427)
--(axis cs:1.39733333333333,9.77542317560427)
--(axis cs:1.39733333333333,38.4034668294243)
--(axis cs:1.136,38.4034668294243)
--(axis cs:1.136,9.77542317560427)
--cycle;
\path [draw=black, fill=color1, line width=0.48pt]
(axis cs:1.60266666666667,22.4372303019783)
--(axis cs:1.864,22.4372303019783)
--(axis cs:1.864,47.779798905554)
--(axis cs:1.60266666666667,47.779798905554)
--(axis cs:1.60266666666667,22.4372303019783)
--cycle;
\path [draw=black, fill=color2, line width=0.48pt]
(axis cs:1.86933333333333,20.3661601748041)
--(axis cs:2.13066666666667,20.3661601748041)
--(axis cs:2.13066666666667,47.0497220362408)
--(axis cs:1.86933333333333,47.0497220362408)
--(axis cs:1.86933333333333,20.3661601748041)
--cycle;
\path [draw=black, fill=color3, line width=0.48pt]
(axis cs:2.136,16.8547858128858)
--(axis cs:2.39733333333333,16.8547858128858)
--(axis cs:2.39733333333333,45.3338866282839)
--(axis cs:2.136,45.3338866282839)
--(axis cs:2.136,16.8547858128858)
--cycle;

\addplot [line width=0.48pt, black, forget plot]
table {%
-0.266666666666667 11.7289755418051
-0.266666666666667 -23.5966114597364
};
\addplot [line width=0.48pt, black, forget plot]
table {%
-0.266666666666667 35.2825253928013
-0.266666666666667 54.8185379658112
};
\addplot [line width=0.48pt, black, forget plot]
table {%
-0.332 -23.5966114597364
-0.201333333333333 -23.5966114597364
};
\addplot [line width=0.48pt, black, forget plot]
table {%
-0.332 54.8185379658112
-0.201333333333333 54.8185379658112
};
\addplot [line width=0.48pt, black, forget plot]
table {%
0 9.11200304474079
0 -28.741810398125
};
\addplot [line width=0.48pt, black, forget plot]
table {%
0 34.3689905730001
0 54.8185379658112
};
\addplot [line width=0.48pt, black, forget plot]
table {%
-0.0653333333333333 -28.741810398125
0.0653333333333333 -28.741810398125
};
\addplot [line width=0.48pt, black, forget plot]
table {%
-0.0653333333333333 54.8185379658112
0.0653333333333333 54.8185379658112
};
\addplot [line width=0.48pt, black, forget plot]
table {%
0.266666666666667 6.29217439327446
0.266666666666667 -33.385208876748
};
\addplot [line width=0.48pt, black, forget plot]
table {%
0.266666666666667 32.7440241294636
0.266666666666667 54.8185379658112
};
\addplot [line width=0.48pt, black, forget plot]
table {%
0.201333333333333 -33.385208876748
0.332 -33.385208876748
};
\addplot [line width=0.48pt, black, forget plot]
table {%
0.201333333333333 54.8185379658112
0.332 54.8185379658112
};
\addplot [line width=0.48pt, black, forget plot]
table {%
0.733333333333333 16.060533220124
0.733333333333333 -21.397209559057
};
\addplot [line width=0.48pt, black, forget plot]
table {%
0.733333333333333 41.0328867512672
0.733333333333333 60.8331067264333
};
\addplot [line width=0.48pt, black, forget plot]
table {%
0.668 -21.397209559057
0.798666666666667 -21.397209559057
};
\addplot [line width=0.48pt, black, forget plot]
table {%
0.668 60.8331067264333
0.798666666666667 60.8331067264333
};
\addplot [line width=0.48pt, black, forget plot]
table {%
1 12.9477614391104
1 -27.8842987526128
};
\addplot [line width=0.48pt, black, forget plot]
table {%
1 40.1701089263648
1 60.8331067264333
};
\addplot [line width=0.48pt, black, forget plot]
table {%
0.934666666666667 -27.8842987526128
1.06533333333333 -27.8842987526128
};
\addplot [line width=0.48pt, black, forget plot]
table {%
0.934666666666667 60.8331067264333
1.06533333333333 60.8331067264333
};
\addplot [line width=0.48pt, black, forget plot]
table {%
1.26666666666667 9.77542317560427
1.26666666666667 -33.1562351886878
};
\addplot [line width=0.48pt, black, forget plot]
table {%
1.26666666666667 38.4034668294243
1.26666666666667 60.8331067264333
};
\addplot [line width=0.48pt, black, forget plot]
table {%
1.20133333333333 -33.1562351886878
1.332 -33.1562351886878
};
\addplot [line width=0.48pt, black, forget plot]
table {%
1.20133333333333 60.8331067264333
1.332 60.8331067264333
};
\addplot [line width=0.48pt, black, forget plot]
table {%
1.73333333333333 22.4372303019783
1.73333333333333 -15.5720541249825
};
\addplot [line width=0.48pt, black, forget plot]
table {%
1.73333333333333 47.779798905554
1.73333333333333 66.8150201390685
};
\addplot [line width=0.48pt, black, forget plot]
table {%
1.668 -15.5720541249825
1.79866666666667 -15.5720541249825
};
\addplot [line width=0.48pt, black, forget plot]
table {%
1.668 66.8150201390685
1.79866666666667 66.8150201390685
};
\addplot [line width=0.48pt, black, forget plot]
table {%
2 20.3661601748041
2 -19.649095261247
};
\addplot [line width=0.48pt, black, forget plot]
table {%
2 47.0497220362408
2 66.8150201390685
};
\addplot [line width=0.48pt, black, forget plot]
table {%
1.93466666666667 -19.649095261247
2.06533333333333 -19.649095261247
};
\addplot [line width=0.48pt, black, forget plot]
table {%
1.93466666666667 66.8150201390685
2.06533333333333 66.8150201390685
};
\addplot [line width=0.48pt, black, forget plot]
table {%
2.26666666666667 16.8547858128858
2.26666666666667 -25.8551115107199
};
\addplot [line width=0.48pt, black, forget plot]
table {%
2.26666666666667 45.3338866282839
2.26666666666667 66.8150201390685
};
\addplot [line width=0.48pt, black, forget plot]
table {%
2.20133333333333 -25.8551115107199
2.332 -25.8551115107199
};
\addplot [line width=0.48pt, black, forget plot]
table {%
2.20133333333333 66.8150201390685
2.332 66.8150201390685
};
\addplot [line width=0.48pt, black, forget plot]
table {%
-0.397333333333333 25.294835169323
-0.136 25.294835169323
};
\addplot [line width=0.48pt, black, forget plot]
table {%
-0.130666666666667 22.8225647626622
0.130666666666667 22.8225647626622
};
\addplot [line width=0.48pt, black, forget plot]
table {%
0.136 19.5937020739884
0.397333333333333 19.5937020739884
};
\addplot [line width=0.48pt, black, forget plot]
table {%
0.602666666666667 30.6542903467482
0.864 30.6542903467482
};
\addplot [line width=0.48pt, black, forget plot]
table {%
0.869333333333333 28.6559468658653
1.13066666666667 28.6559468658653
};
\addplot [line width=0.48pt, black, forget plot]
table {%
1.136 25.2213902184372
1.39733333333333 25.2213902184372
};
\addplot [line width=0.48pt, black, forget plot]
table {%
1.60266666666667 37.4402741561415
1.864 37.4402741561415
};
\addplot [line width=0.48pt, black, forget plot]
table {%
1.86933333333333 35.6442914226256
2.13066666666667 35.6442914226256
};
\addplot [line width=0.48pt, black, forget plot]
table {%
2.136 32.4231448876694
2.39733333333333 32.4231448876694
};
\end{axis}

\end{tikzpicture}
    \end{subfigure}
    \caption{\gls{sinr} vs. $n_{tx}$ in the 2W-LD scenario, with PRFS-1 and $D=20$~Mbps.}
    \label{fig:sci}
\end{figure}
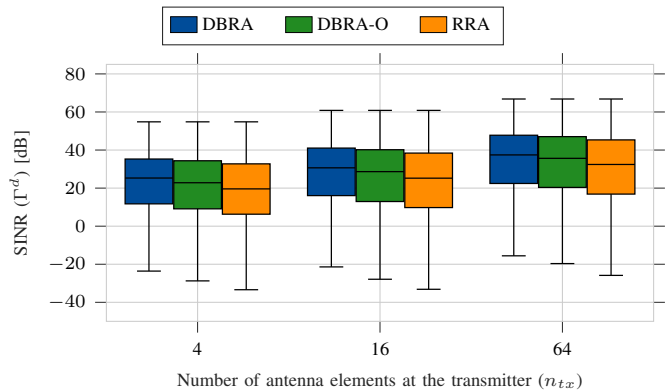

\begin{figure*}[t!]
        \begin{subfigure}[b]{\linewidth}
	\centering
%
%

\definecolor{black25}{RGB}{25,25,25}
\definecolor{mediumaquamarine102194165}{RGB}{102,194,165}
\definecolor{limegreen3122331}{RGB}{31,223,31}
\definecolor{silver}{RGB}{192,192,192}
\definecolor{lightgreen178223138}{RGB}{178,223,138}
\definecolor{lightsteelblue173203219}{RGB}{173,203,219}
\definecolor{steelblue31120180}{RGB}{31,120,180}

\definecolor{color1}{RGB}{0, 76, 153}
\definecolor{color2}{RGB}{34, 139, 34}
\definecolor{color3}{RGB}{255, 140, 0}

\begin{tikzpicture}
\pgfplotsset{every tick label/.append style={font=\scriptsize}}

\pgfplotsset{compat=1.11,
	/pgfplots/ybar legend/.style={
		/pgfplots/legend image code/.code={%
			\draw[##1,/tikz/.cd,yshift=-0.25em]
			(0cm,0cm) rectangle (10pt,0.6em);},
	},
}

\begin{axis}[%
width=0,
height=0,
at={(0,0)},
scale only axis,
xmin=0,
xmax=0,
xtick={},
ymin=0,
ymax=0,
ytick={},
axis background/.style={fill=white},
legend style={legend cell align=left,
              align=center,
              draw=white!15!black,
              at={(0.5, 1.3)},
              anchor=center,
              /tikz/every even column/.append style={column sep=1em}},
legend columns=3,
]
\addplot[ybar,ybar legend,draw=black,fill=color1,line width=0.08pt]
table[row sep=crcr]{%
	0	0\\
};
\addlegendentry{DBRA}

\addplot[ybar legend,ybar,draw=black,fill=color2,line width=0.08pt]
  table[row sep=crcr]{%
	0	0\\
};
\addlegendentry{DBRA-O}

\addplot[ybar legend,ybar,draw=black,fill=color3,line width=0.08pt]
  table[row sep=crcr]{%
	0	0\\
};
\addlegendentry{RRA}

\end{axis}
\end{tikzpicture}%
	\end{subfigure}
 \vskip 0.1cm
    \centering
   \subfloat[][$n_{tx}= 4$.]
	{
\begin{tikzpicture}

\definecolor{darkgray176}{RGB}{176,176,176}
\definecolor{darkslategray38}{RGB}{38,38,38}
\definecolor{lightgray204}{RGB}{204,204,204}
\definecolor{color1}{RGB}{0, 76, 153}
\definecolor{color2}{RGB}{34, 139, 34}
\definecolor{color3}{RGB}{255, 140, 0}

\begin{axis}[
width = \textwidth/2.1,
height = 5cm,
tick align=outside,
axis line style={lightgray204},
x grid style={lightgray204},
xlabel={Scenario},
xmajorgrids,
xmin=-0.5, xmax=3.5,
xtick style={darkslategray38},
xtick={0,1,2,3},
xticklabels={1W-LD,1W-HD,2W-LD,2W-HD},
y grid style={darkgray176},
ylabel={Average collision probability [\%]},
ymajorgrids,
ymin=0.1, ymax=20.1,
ytick={0,5,10,15,20},
ytick style={darkslategray38}
]
\draw[draw=black,fill=color1,line width=0.32pt] (axis cs:-0.4,0) rectangle (axis cs:-0.133333333333333,2.18542550260663);

\draw[draw=black,fill=color1,line width=0.32pt] (axis cs:0.6,0) rectangle (axis cs:0.866666666666667,11.3583425223587);
\draw[draw=black,fill=color1,line width=0.32pt] (axis cs:1.6,0) rectangle (axis cs:1.86666666666667,3.73585158762105);
\draw[draw=black,fill=color1,line width=0.32pt] (axis cs:2.6,0) rectangle (axis cs:2.86666666666667,13.3623320005948);
\draw[draw=black,fill=color2,line width=0.32pt] (axis cs:-0.133333333333333,0) rectangle (axis cs:0.133333333333333,3.13029634366096);

\draw[draw=black,fill=color2,line width=0.32pt] (axis cs:0.866666666666667,0) rectangle (axis cs:1.13333333333333,14.6487708289752);
\draw[draw=black,fill=color2,line width=0.32pt] (axis cs:1.86666666666667,0) rectangle (axis cs:2.13333333333333,5.7092518992652);
\draw[draw=black,fill=color2,line width=0.32pt] (axis cs:2.86666666666667,0) rectangle (axis cs:3.13333333333333,15.5393048601714);
\draw[draw=black,fill=color3,line width=0.32pt] (axis cs:0.133333333333333,0) rectangle (axis cs:0.4,4.44292827915722);

\draw[draw=black,fill=color3,line width=0.32pt] (axis cs:1.13333333333333,0) rectangle (axis cs:1.4,18.5443120313795);
\draw[draw=black,fill=color3,line width=0.32pt] (axis cs:2.13333333333333,0) rectangle (axis cs:2.4,7.96104022847837);
\draw[draw=black,fill=color3,line width=0.32pt] (axis cs:3.13333333333333,0) rectangle (axis cs:3.4,18.8843048468946);
\end{axis}

\end{tikzpicture}
            \label{fig:sci_4_2}
	}
   \subfloat[][$n_{tx}= 64$.]
	{
\begin{tikzpicture}

\definecolor{darkgray176}{RGB}{176,176,176}
\definecolor{darkslategray38}{RGB}{38,38,38}
\definecolor{lightgray204}{RGB}{204,204,204}
\definecolor{color1}{RGB}{0, 76, 153}
\definecolor{color2}{RGB}{34, 139, 34}
\definecolor{color3}{RGB}{255, 140, 0}

\begin{axis}[
width = \textwidth/2.1,
height = 5cm,
axis line style={lightgray204},
tick align=outside,
x grid style={lightgray204},
xlabel={Scenario},
xmajorgrids,
xmin=-0.5, xmax=3.5,
xtick style={darkslategray38},
xtick={0,1,2,3},
xticklabels={1W-LD,1W-HD,2W-LD,2W-HD},
y grid style={darkgray176},
ylabel={Average collision probability [\%]},
ymajorgrids,
ymin=0.1, ymax=20.1,
ytick={0,5,10,15,20},
ytick style={darkslategray38},
]

\draw[draw=black,fill=color1,line width=0.32pt] (axis cs:-0.4,0) rectangle (axis cs:-0.133333333333333,1.15674002980088);

\draw[draw=black,fill=color1,line width=0.32pt] (axis cs:0.6,0) rectangle (axis cs:0.866666666666667,7.43276395732111);
\draw[draw=black,fill=color1,line width=0.32pt] (axis cs:1.6,0) rectangle (axis cs:1.86666666666667,2.10397654598958);
\draw[draw=black,fill=color1,line width=0.32pt] (axis cs:2.6,0) rectangle (axis cs:2.86666666666667,7.17561616747071);
\draw[draw=black,fill=color2,line width=0.32pt] (axis cs:-0.133333333333333,0) rectangle (axis cs:0.133333333333333,1.63486475027027);

\draw[draw=black,fill=color2,line width=0.32pt] (axis cs:0.866666666666667,0) rectangle (axis cs:1.13333333333333,9.32818170816749);
\draw[draw=black,fill=color2,line width=0.32pt] (axis cs:1.86666666666667,0) rectangle (axis cs:2.13333333333333,2.76372151309315);
\draw[draw=black,fill=color2,line width=0.32pt] (axis cs:2.86666666666667,0) rectangle (axis cs:3.13333333333333,8.34049320957372);
\draw[draw=black,fill=color3,line width=0.32pt] (axis cs:0.133333333333333,0) rectangle (axis cs:0.4,2.8126266902861);

\draw[draw=black,fill=color3,line width=0.32pt] (axis cs:1.13333333333333,0) rectangle (axis cs:1.4,12.1834986543573);
\draw[draw=black,fill=color3,line width=0.32pt] (axis cs:2.13333333333333,0) rectangle (axis cs:2.4,4.05876028117887);
\draw[draw=black,fill=color3,line width=0.32pt] (axis cs:3.13333333333333,0) rectangle (axis cs:3.4,10.5616545077704);
\end{axis}

\end{tikzpicture}
            \label{fig:sci_64_2}
	}
    \caption{Average collision probability in different scenarios vs. $n_{tx}$, with PRFS-1 and $D=20$~Mbps.}
    \label{fig:sci_3}
\end{figure*}
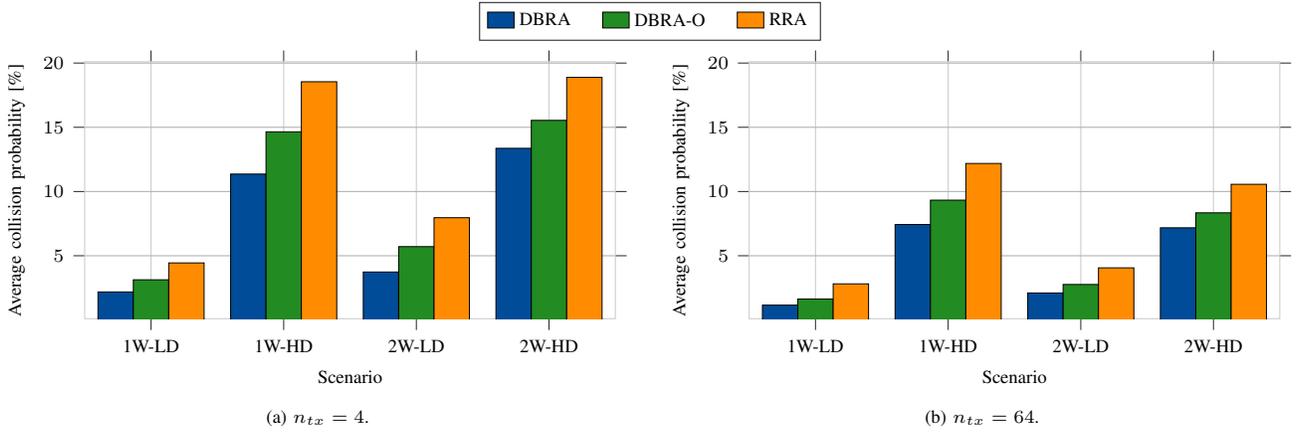

\begin{figure*}[t!]
        \begin{subfigure}[b]{\linewidth}
	\centering
%
%

\definecolor{black25}{RGB}{25,25,25}
\definecolor{mediumaquamarine102194165}{RGB}{102,194,165}
\definecolor{limegreen3122331}{RGB}{31,223,31}
\definecolor{silver}{RGB}{192,192,192}
\definecolor{lightgreen178223138}{RGB}{178,223,138}
\definecolor{lightsteelblue173203219}{RGB}{173,203,219}
\definecolor{steelblue31120180}{RGB}{31,120,180}

\definecolor{color1}{RGB}{0, 76, 153}
\definecolor{color2}{RGB}{34, 139, 34}
\definecolor{color3}{RGB}{255, 140, 0}

\begin{tikzpicture}
\pgfplotsset{every tick label/.append style={font=\scriptsize}}

\pgfplotsset{compat=1.11,
	/pgfplots/ybar legend/.style={
		/pgfplots/legend image code/.code={%
			\draw[##1,/tikz/.cd,yshift=-0.25em]
			(0cm,0cm) rectangle (10pt,0.6em);},
	},
}

\begin{axis}[%
width=0,
height=0,
at={(0,0)},
scale only axis,
xmin=0,
xmax=0,
xtick={},
ymin=0,
ymax=0,
ytick={},
axis background/.style={fill=white},
legend style={legend cell align=left,
              align=center,
              draw=white!15!black,
              at={(0.5, 1.3)},
              anchor=center,
              /tikz/every even column/.append style={column sep=1em}},
legend columns=3,
]
\addplot[ybar,ybar legend,draw=black,fill=color1,line width=0.08pt]
table[row sep=crcr]{%
	0	0\\
};
\addlegendentry{DBRA}

\addplot[ybar legend,ybar,draw=black,fill=color2,line width=0.08pt]
  table[row sep=crcr]{%
	0	0\\
};
\addlegendentry{DBRA-O}

\addplot[ybar legend,ybar,draw=black,fill=color3,line width=0.08pt]
  table[row sep=crcr]{%
	0	0\\
};
\addlegendentry{RRA}

\end{axis}
\end{tikzpicture}%
	\end{subfigure}
 \vskip 0.1cm
    \centering
    \subfloat[][$D=5.7$~Mbps.]
	{
	    \label{fig:cp_5}
\begin{tikzpicture}

\definecolor{darkgray176}{RGB}{176,176,176}
\definecolor{lightgray204}{RGB}{204,204,204}
\definecolor{color1}{RGB}{0, 76, 153}
\definecolor{color2}{RGB}{34, 139, 34}
\definecolor{color3}{RGB}{255, 140, 0}

\begin{axis}[
width = \textwidth/3.1,
height = 5cm,
axis line style={lightgray204},
x grid style={lightgray204},
tick align=outside,
xlabel={Resource frame structure},
xmajorgrids,
xmin=-0.5, xmax=3.5,
xtick style={color=black},
xtick={0,1,2,3},
xticklabels={PRFS-1, PRFS-2, PRFS-3, PRFS-4},
y grid style={darkgray176},
ylabel={Average collision probability [\%]},
ymajorgrids,
ymin=0, ymax=5.22524835776524,
ytick style={color=black}
]
\draw[draw=black,fill=color1,line width=0.32pt] (axis cs:-0.4,0) rectangle (axis cs:-0.133333333333333,1.12456574338007);

\draw[draw=black,fill=color1,line width=0.32pt] (axis cs:0.6,0) rectangle (axis cs:0.866666666666667,1.26446935666873);
\draw[draw=black,fill=color1,line width=0.32pt] (axis cs:1.6,0) rectangle (axis cs:1.86666666666667,1.36959053008813);
\draw[draw=black,fill=color1,line width=0.32pt] (axis cs:2.6,0) rectangle (axis cs:2.86666666666667,1.51607665960607);
\draw[draw=black,fill=color2,line width=0.32pt] (axis cs:-0.133333333333333,0) rectangle (axis cs:0.133333333333333,2.36705215040905);

\draw[draw=black,fill=color2,line width=0.32pt] (axis cs:0.866666666666667,0) rectangle (axis cs:1.13333333333333,2.63866618564316);
\draw[draw=black,fill=color2,line width=0.32pt] (axis cs:1.86666666666667,0) rectangle (axis cs:2.13333333333333,2.94353035130198);
\draw[draw=black,fill=color2,line width=0.32pt] (axis cs:2.86666666666667,0) rectangle (axis cs:3.13333333333333,3.28769011660791);
\draw[draw=black,fill=color3,line width=0.32pt] (axis cs:0.133333333333333,0) rectangle (axis cs:0.4,3.60584597667883);

\draw[draw=black,fill=color3,line width=0.32pt] (axis cs:1.13333333333333,0) rectangle (axis cs:1.4,3.82496261611481);
\draw[draw=black,fill=color3,line width=0.32pt] (axis cs:2.13333333333333,0) rectangle (axis cs:2.4,4.30774552516914);
\draw[draw=black,fill=color3,line width=0.32pt] (axis cs:3.13333333333333,0) rectangle (axis cs:3.4,4.76136075257567);
\end{axis}

\end{tikzpicture}
	}
   \subfloat[][$D=20$~Mbps.]
	{
		\label{fig:cp_20}
\begin{tikzpicture}

\definecolor{darkgray176}{RGB}{176,176,176}
\definecolor{lightgray204}{RGB}{204,204,204}
\definecolor{color1}{RGB}{0, 76, 153}
\definecolor{color2}{RGB}{34, 139, 34}
\definecolor{color3}{RGB}{255, 140, 0}

\begin{axis}[
width = \textwidth/3.1,
height = 5cm,
axis line style={lightgray204},
x grid style={lightgray204},
tick align=outside,
xlabel={Resource frame structure},
xmajorgrids,
xmin=-0.5, xmax=3.5,
xtick style={color=black},
xtick={0,1,2,3},
xticklabels={PRFS-1, PRFS-2, PRFS-3, PRFS-4},
y grid style={darkgray176},
ylabel={Average collision probability [\%]},
ymajorgrids,
ymin=0, ymax=16.6952057594046,
ytick style={color=black}
]

\draw[draw=black,fill=color1,line width=0.32pt] (axis cs:-0.4,0) rectangle (axis cs:-0.133333333333333,7.43276395732111);

\draw[draw=black,fill=color1,line width=0.32pt] (axis cs:0.6,0) rectangle (axis cs:0.866666666666667,8.99266819457283);
\draw[draw=black,fill=color1,line width=0.32pt] (axis cs:1.6,0) rectangle (axis cs:1.86666666666667,8.71426110986384);
\draw[draw=black,fill=color1,line width=0.32pt] (axis cs:2.6,0) rectangle (axis cs:2.86666666666667,10.6909591030487);
\draw[draw=black,fill=color2,line width=0.32pt] (axis cs:-0.133333333333333,0) rectangle (axis cs:0.133333333333333,9.32818170816749);

\draw[draw=black,fill=color2,line width=0.32pt] (axis cs:0.866666666666667,0) rectangle (axis cs:1.13333333333333,10.243685972659);
\draw[draw=black,fill=color2,line width=0.32pt] (axis cs:1.86666666666667,0) rectangle (axis cs:2.13333333333333,11.0090287504287);
\draw[draw=black,fill=color2,line width=0.32pt] (axis cs:2.86666666666667,0) rectangle (axis cs:3.13333333333333,12.4518243582761);
\draw[draw=black,fill=color3,line width=0.32pt] (axis cs:0.133333333333333,0) rectangle (axis cs:0.4,12.1834986543573);

\draw[draw=black,fill=color3,line width=0.32pt] (axis cs:1.13333333333333,0) rectangle (axis cs:1.4,12.3130166478966);
\draw[draw=black,fill=color3,line width=0.32pt] (axis cs:2.13333333333333,0) rectangle (axis cs:2.4,13.4542019775518);
\draw[draw=black,fill=color3,line width=0.32pt] (axis cs:3.13333333333333,0) rectangle (axis cs:3.4,14.9434986768183);
\end{axis}

\end{tikzpicture}
	}
   \subfloat[][$D=40$~Mbps.]
	{
		\label{fig:cp_40}
\begin{tikzpicture}

\definecolor{darkgray176}{RGB}{176,176,176}
\definecolor{lightgray204}{RGB}{204,204,204}
\definecolor{color1}{RGB}{0, 76, 153}
\definecolor{color2}{RGB}{34, 139, 34}
\definecolor{color3}{RGB}{255, 140, 0}

\begin{axis}[
width = \textwidth/3.1,
height = 5cm,
axis line style={lightgray204},
x grid style={lightgray204},
tick align=outside,
x grid style={darkgray176},
xlabel={Resource frame structure},
xmajorgrids,
xmin=-0.5, xmax=3.5,
xtick style={color=black},
xtick={0,1,2,3},
xticklabels={PRFS-1, PRFS-2, PRFS-3, PRFS-4},
y grid style={darkgray176},
ylabel={Average collision probability [\%]},
ymajorgrids,
ymin=0, ymax=29.3186629307938,
ytick style={color=black}
]
\draw[draw=black,fill=color1,line width=0.32pt] (axis cs:-0.4,0) rectangle (axis cs:-0.133333333333333,18.0465488216216);

\draw[draw=black,fill=color1,line width=0.32pt] (axis cs:0.6,0) rectangle (axis cs:0.866666666666667,21.9739936664211);
\draw[draw=black,fill=color1,line width=0.32pt] (axis cs:1.6,0) rectangle (axis cs:1.86666666666667,22.5600786445883);
\draw[draw=black,fill=color1,line width=0.32pt] (axis cs:2.6,0) rectangle (axis cs:2.86666666666667,25.8611228569761);
\draw[draw=black,fill=color2,line width=0.32pt] (axis cs:-0.133333333333333,0) rectangle (axis cs:0.133333333333333,18.4494441771135);

\draw[draw=black,fill=color2,line width=0.32pt] (axis cs:0.866666666666667,0) rectangle (axis cs:1.13333333333333,21.8893159560457);
\draw[draw=black,fill=color2,line width=0.32pt] (axis cs:1.86666666666667,0) rectangle (axis cs:2.13333333333333,23.0615920806889);
\draw[draw=black,fill=color2,line width=0.32pt] (axis cs:2.86666666666667,0) rectangle (axis cs:3.13333333333333,25.808718362875);
\draw[draw=black,fill=color3,line width=0.32pt] (axis cs:0.133333333333333,0) rectangle (axis cs:0.4,19.4395231573499);

\draw[draw=black,fill=color3,line width=0.32pt] (axis cs:1.13333333333333,0) rectangle (axis cs:1.4,21.936474074297);
\draw[draw=black,fill=color3,line width=0.32pt] (axis cs:2.13333333333333,0) rectangle (axis cs:2.4,23.8547833338722);
\draw[draw=black,fill=color3,line width=0.32pt] (axis cs:3.13333333333333,0) rectangle (axis cs:3.4,26.0212061850248);
\end{axis}

\end{tikzpicture}
	}
    \caption{Average collision probability vs. $D$ and PRFS in the 1W-HD scenario, with $n_{tx}=64$.}    
    \label{fig:collision}
\end{figure*}
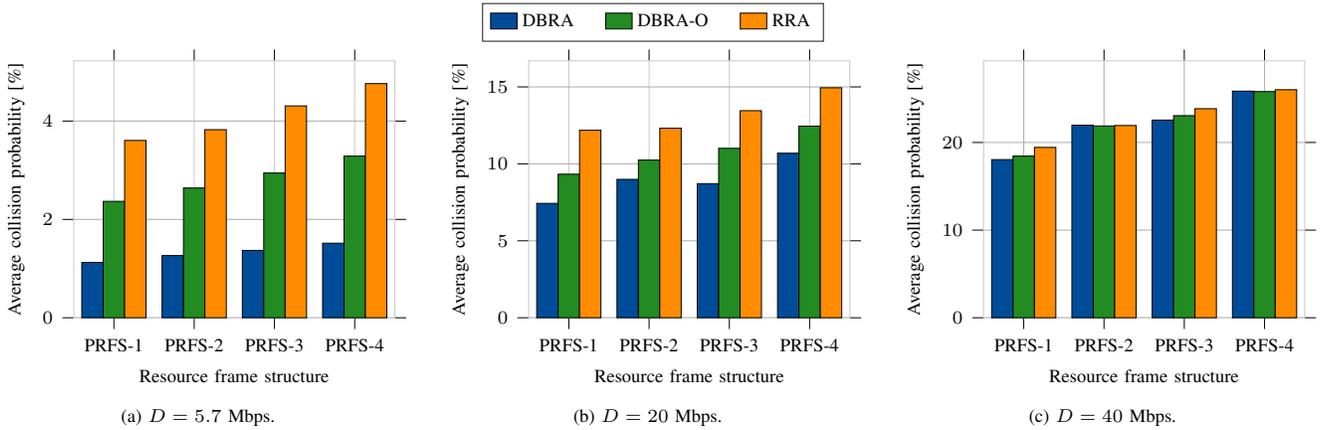

\subsection{Simulation Results}
\label{sub:results}

\begin{table}[t!]
    \centering
    \caption{Vehicle statistics in the considered urban scenarios.}
    \label{tab:vehicle_distances}
    \begin{tabular}{lccccc}
        \toprule
        \textbf{Scenario} & \textbf{Vehicles} & \textbf{Avg. distance [m]} & \textbf{Std. dev. distance [m]} \\
        \midrule
        1W-LD  & 15  & 54.2  & 42.0  \\
        1W-HD & 30  & 19.2   & 17.4  \\
        2W-LD  & 30  & 64.5 & 42.1  \\
        2W-HD & 60  & 21.0  & 19.7  \\
        \bottomrule
    \end{tabular}
    \vskip -0.2cm
\end{table}

\begin{table}[t!]
    \centering
    \vspace{0.3cm}
    \caption{ Number of resources to be allocated for SCI transmissions per slot in the primary and paired directions.}
    \begin{tabular}{@{\hskip 10pt}l@{\hskip 20pt}l@{\hskip 20pt}l@{\hskip 10pt}}  
        \toprule
        \textbf{Scheme}  & \textbf{Primary Direction} & \textbf{Paired Direction} \\ \midrule
        DBRA & 12 PRBs + 48 bits & 12 PRBs  \\
        DBRA-O & 12 PRBs  & N/A  \\
        RRA & N/A & N/A  \\ \bottomrule
    \end{tabular}
    \label{tab:resource_allocation}
\end{table}

In the following results, we assume that each vehicle can communicate with $K_{rx}=5$ neighbors.
\cref{fig:sci} shows the boxplot of the \gls{sinr} for the three resource allocation schemes.
We observe that DBRA consistently outperforms its competitors in terms of both median and minimum \gls{sinr}.
In fact, DBRA can sense \gls{sci} in both primary and paired directions, thereby improving resource selection. As expected, as $n_{tx}$ increases, the \gls{sinr} also increases given the higher beamforming gain.

As far as the antenna configuration is concerned, in \cref{fig:sci_3} we observe that the collision probability decreases as $n_{tx}$ increases: for instance, in the 2W-HD scenario, it is approximately 13\% for $n_{tx}=4$ vs. 7\% for $n_{tx}=64$ for DBRA.
On one side, $n_{tx}=4$ configures larger beams and the vehicles can receive more \gls{sci} transmissions, reducing the exposed node problem and resulting in better resource allocation. On the other side, $n_{tx}=64$ provides more spatial diversity by beamforming, which also improves the SINR.
Overall, the latter is the dominant component, which demonstrates that directional resource allocation is feasible and convenient in NR V2X.
In general, DBRA outperforms DBRA-O given that more SCI transmissions are received: in the 1W-HD scenario and for $n_{tx}=64$, the collision probability goes from 11\% to almost 15\%, respectively.
Notice that the collision probability is higher in 2W-HD and 1W-HD  compared to 2W-LD and 1W-LD since the network is more congested (i.e., there are more vehicles).

Since the trends for the average collision probability in the different scenarios are similar, we focus on the 1W-HD scenario for the remainder of our evaluation.
\cref{fig:collision} illustrates the average collision probability as a function of the \gls{prfs} configuration and the application rate.
We observe that the collision performance improves as \gls{prfs} decreases for all resource allocation schemes, and it is up to 25\% for PRFS-4 and $D=40$ Mbps.
As the numerology decreases, the PRB size also decreases, which effectively reduces the number of resources that need to be allocated for sending data.
As such, the system configures fewer transmissions, so the collision probability decreases.
At the same time, with a smaller (narrower) PRB, the impact of thermal noise is less significant, which improves the SINR and further demonstrates the lower collision probability as PRFS~decreases.

\begin{figure*}[t!]
    \begin{subfigure}[b]{\linewidth}
    \centering
%
%

\definecolor{black25}{RGB}{25,25,25}
\definecolor{mediumaquamarine102194165}{RGB}{102,194,165}
\definecolor{limegreen3122331}{RGB}{31,223,31}
\definecolor{silver}{RGB}{192,192,192}
\definecolor{lightgreen178223138}{RGB}{178,223,138}
\definecolor{lightsteelblue173203219}{RGB}{173,203,219}
\definecolor{steelblue31120180}{RGB}{31,120,180}

\definecolor{color1}{RGB}{0, 76, 153}
\definecolor{color2}{RGB}{34, 139, 34}
\definecolor{color3}{RGB}{255, 140, 0}

\begin{tikzpicture}
\pgfplotsset{every tick label/.append style={font=\scriptsize}}

\pgfplotsset{compat=1.11,
	/pgfplots/ybar legend/.style={
		/pgfplots/legend image code/.code={%
			\draw[##1,/tikz/.cd,yshift=-0.25em]
			(0cm,0cm) rectangle (10pt,0.6em);},
	},
}

\begin{axis}[%
width=0,
height=0,
at={(0,0)},
scale only axis,
xmin=0,
xmax=0,
xtick={},
ymin=0,
ymax=0,
ytick={},
axis background/.style={fill=white},
legend style={legend cell align=left,
              align=center,
              draw=white!15!black,
              at={(0.5, 1.3)},
              anchor=center,
              /tikz/every even column/.append style={column sep=1em}},
legend columns=3,
]
\addplot[ybar,ybar legend,draw=black,fill=color1,line width=0.08pt]
table[row sep=crcr]{%
	0	0\\
};
\addlegendentry{DBRA}

\addplot[ybar legend,ybar,draw=black,fill=color2,line width=0.08pt]
  table[row sep=crcr]{%
	0	0\\
};
\addlegendentry{DBRA-O}

\addplot[ybar legend,ybar,draw=black,fill=color3,line width=0.08pt]
  table[row sep=crcr]{%
	0	0\\
};
\addlegendentry{RRA}

\end{axis}
\end{tikzpicture}%
    \end{subfigure}
    \vskip 0.1cm
    \centering
    \subfloat[][PDB $=10$~ms.]
	{
	    \label{fig:sl_antenna}
\begin{tikzpicture}

\definecolor{darkgray176}{RGB}{176,176,176}
\definecolor{darkslategray38}{RGB}{38,38,38}
\definecolor{lightgray204}{RGB}{204,204,204}
\definecolor{color1}{RGB}{0, 76, 153}
\definecolor{color2}{RGB}{34, 139, 34}
\definecolor{color3}{RGB}{255, 140, 0}

\begin{axis}[
width = \textwidth/2.1,
height = 5cm,
axis line style={lightgray204},
x grid style={lightgray204},
tick pos=both,
x grid style={darkgray176},
xlabel=\textcolor{darkslategray38}{Number of TX antenna elements ($n_{tx}$)},
xmajorgrids,
xmin=-0.5, xmax=2.5,
xtick style={color=black},
xtick={0,1,2},
xticklabels={4, 16, 64},
y grid style={darkgray176},
ylabel={Average collision probability [\%]},
ymajorgrids,
ymin=0, ymax=3.91143038423899,
ytick style={color=black}
]
\draw[draw=black,fill=color1, line width=0.32pt] (axis cs:-0.4,0) rectangle (axis cs:-0.133333333333333,2.98986976954263);

\draw[draw=black,fill=color1, line width=0.32pt] (axis cs:0.6,0) rectangle (axis cs:0.866666666666667,2.21047650797406);
\draw[draw=black,fill=color1, line width=0.32pt] (axis cs:1.6,0) rectangle (axis cs:1.86666666666667,1.59293034660013);
\draw[draw=black,fill=color2, line width=0.32pt] (axis cs:-0.133333333333333,0) rectangle (axis cs:0.133333333333333,3.20306702943358);

\draw[draw=black,fill=color2, line width=0.32pt] (axis cs:0.866666666666667,0) rectangle (axis cs:1.13333333333333,2.60272720429298);
\draw[draw=black,fill=color2, line width=0.32pt] (axis cs:1.86666666666667,0) rectangle (axis cs:2.13333333333333,1.69428642097452);
\draw[draw=black,fill=color3, line width=0.32pt] (axis cs:0.133333333333333,0) rectangle (axis cs:0.4,3.72517179451332);

\draw[draw=black,fill=color3, line width=0.32pt] (axis cs:1.13333333333333,0) rectangle (axis cs:1.4,2.84711094860682);
\draw[draw=black,fill=color3, line width=0.32pt] (axis cs:2.13333333333333,0) rectangle (axis cs:2.4,1.97469593177687);
\end{axis}

\end{tikzpicture}
	}
   \subfloat[][$n_{tx}=64$.]
	{
		\label{fig:sl_pdb}
\begin{tikzpicture}

\definecolor{darkgray176}{RGB}{176,176,176}
\definecolor{lightgray204}{RGB}{204,204,204}
\definecolor{color1}{RGB}{0, 76, 153}
\definecolor{color2}{RGB}{34, 139, 34}
\definecolor{color3}{RGB}{255, 140, 0}

\begin{axis}[
width = \textwidth/2.1,
height = 5cm,
axis line style={lightgray204},
x grid style={lightgray204},
tick pos=both,
x grid style={darkgray176},
xlabel={PDB [ms]},
xmajorgrids,
xmin=-0.5, xmax=2.5,
xtick style={color=black},
xtick={0,1,2},
xticklabels={10,25,50},
y grid style={darkgray176},
ylabel={Average collision probability [\%]},
ymajorgrids,
ymin=0, ymax=3.47839529406704,
ytick style={color=black}
]
\draw[draw=black,fill=color1,line width=0.32pt] (axis cs:-0.4,0) rectangle (axis cs:-0.133333333333333,1.59293034660013);

\draw[draw=black,fill=color1,line width=0.32pt] (axis cs:0.6,0) rectangle (axis cs:0.866666666666667,1.28993751949155);
\draw[draw=black,fill=color1,line width=0.32pt] (axis cs:1.6,0) rectangle (axis cs:1.86666666666667,1.05469464130165);
\draw[draw=black,fill=color2,line width=0.32pt] (axis cs:-0.133333333333333,0) rectangle (axis cs:0.133333333333333,1.69428642097452);

\draw[draw=black,fill=color2,line width=0.32pt] (axis cs:0.866666666666667,0) rectangle (axis cs:1.13333333333333,2.10239920850853);
\draw[draw=black,fill=color2,line width=0.32pt] (axis cs:1.86666666666667,0) rectangle (axis cs:2.13333333333333,2.23655486132768);
\draw[draw=black,fill=color3,line width=0.32pt] (axis cs:0.133333333333333,0) rectangle (axis cs:0.4,1.97469593177687);

\draw[draw=black,fill=color3,line width=0.32pt] (axis cs:1.13333333333333,0) rectangle (axis cs:1.4,3.13612362752583);
\draw[draw=black,fill=color3,line width=0.32pt] (axis cs:2.13333333333333,0) rectangle (axis cs:2.4,3.31275742292099);
\end{axis}

\end{tikzpicture}
	}
    \caption{Average collision probability vs. PDB and $n_{tx}$ in the 1W-HD scenario, with $K_{rx}=1$ and $D=5.7$ Mbps.\vspace{-0.5cm}}
    \label{fig:singlelink}
\end{figure*}
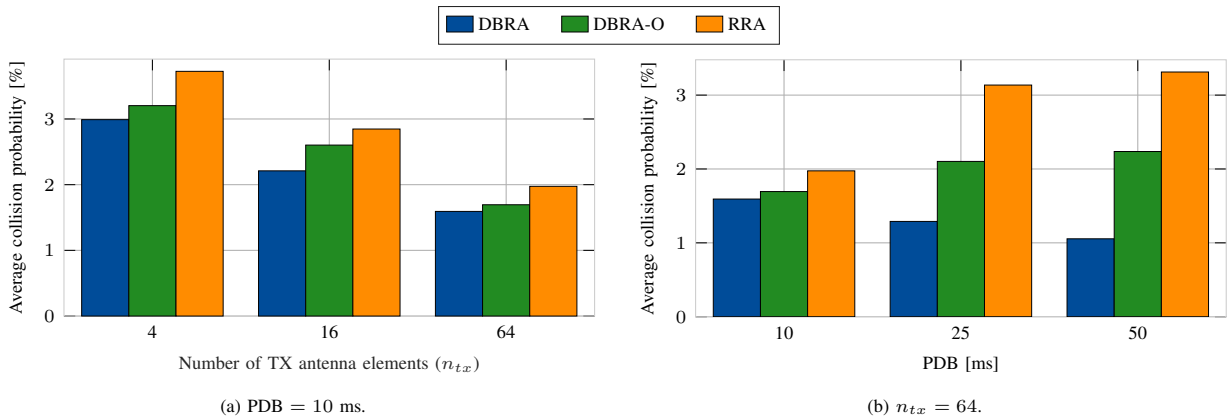

Another important observation is that, as the application rate increases, the average collision probability also increases. 
This is because vehicles are required to use more resources (i.e., select more \glspl{prb} in the Selection Window) to transmit data within the \gls{pdb}.
Interestingly, as $D$ increases, the benefits of DBRA with respect to its competitors are less evident since vehicles tend to use all the available resources, regardless of the optimization approach. 
When $D=5.7$ and $20$~Mbps, the total number of available resources is likely higher than the number of resources to be allocated for sending data, so we can optimize the selection of the available resources based on the \gls{sci}: 
for DBRA, the collision probability is up to 50\% and 70\% lower than DBRA-O and RRA, respectively.
In contrast, for $D=40$~Mbps the system is close to the saturation point, and the collision probability becomes comparable in all schemes.
For example, for PRFS-1, $s_{s}=50\,352$ according to~\cref{eq:s_tb}, and we have 80 slots of $125$ $\mu$s within a 10-ms \gls{pdb}.
Then, for $D=5.7$ ($s_d=0.57$ Mb), we need at least $N_s=\ceil{s_d/s_{s}}=12$ slots for data transmission, which is only 15\% of the available resources.
For $D=40$ Mbps ($s_d=4$ Mb), we have $N_{s}\simeq80$, meaning that 100\% of the resources are consumed.



Now, we consider another scenario where each vehicle only communicates with one receiver (i.e., $K_{rx}=1$).
In \cref{fig:singlelink} we evaluate the impact of the \gls{pdb} and the antenna array size when $D=5.7$~Mbps.
Notably,
as the PDB decreases, the collision probability of RRA improves, while for DBRA it is the opposite. This is due to the fact that, as we set more critical latency constraints, the total number of resources in the Selection Window decreases, and vehicles may be required to use most of the resources to satisfy the PDB requirement, regardless of the available SCI.
Consequently, DBRA tends to approximate RRA.

\section{Conclusions}\label{sec:conclusion}
In this work, we explored uncoordinated resource allocation for NR V2X Mode 2 in the 60 GHz unlicensed bands. In this scheme, vehicles independently select resources for transmission from a predefined pool, using the SCI received from their neighbors to facilitate this selection. Specifically, the SCI is used to identify the available resources, and reduce the probability of collision. While NR V2X assumes omnidirectional sensing, we considered
directional beamforming for both data and SCI transmissions. We evaluated the performance of this approach via simulations as a function of the density of vehicles, the \gls{prfs}, the antenna array size, and the \gls{pdb}. We demonstrated that directional resource allocation via SCI can actually reduce the probability of collision compared to some benchmark schemes.

As part of our future work, we will consider the impact of mobility, as well as errors in the estimate of the position of the vehicles. Moreover, we will formulate an analytical model for the collision probability to validate our simulation~results.

\bibliographystyle{IEEEtran}
\bibliography{bibliography.bib}

\end{document}